\documentclass[12pt,preprint]{aastex}

\newcommand{\Msun}{M$_\odot$}
\newcommand{\Nifs}{$^{56}$Ni}
\newcommand{\Cofs}{$^{56}$Co}
\newcommand{\Fefs}{$^{56}$Fe}

\shorttitle{SN 2003bg}
\shortauthors{Hamuy et al.}

\begin{document}

\title{Supernova 2003bg: The First Type IIb Hypernova\footnote{This 
paper includes data gathered with the 6.5-m
Magellan telescopes located at Las Campanas Observatory, Chile, and
is partly based on observations collected at the Cerro Tololo
Inter-American Observatory, operated by the Association of
Universities for Research in Astronomy, Inc. (AURA), under
cooperative agreement with the National Science Foundation.} }

\author{Mario Hamuy\altaffilmark{2}, Jinsong Deng\altaffilmark{3}, Paolo A. Mazzali\altaffilmark{4}, Nidia I. Morrell\altaffilmark{5}, Mark M. Phillips\altaffilmark{5}, Miguel Roth\altaffilmark{5}, Sergio Gonzalez\altaffilmark{5}, Joanna Thomas-Osip\altaffilmark{5}, Wojtek Krzeminski\altaffilmark{5}, Carlos Contreras\altaffilmark{5}, Jos\'e Maza\altaffilmark{2}, Luis Gonz\'alez\altaffilmark{2}, Leonor Huerta\altaffilmark{2}, Gast\'on Folatelli\altaffilmark{2}, Ryan Chornock\altaffilmark{6}, Alexei V. Filippenko\altaffilmark{6}, S. E. Persson\altaffilmark{7}, W. L. Freedman\altaffilmark{7}, Kathleen Koviak\altaffilmark{7}, Nicholas B. Suntzeff\altaffilmark{8}, and Kevin Krisciunas\altaffilmark{8}}

\altaffiltext{2}{Departamento de Astronom\'\i a, Universidad de Chile, Casilla 36-D, Santiago, Chile.}
\altaffiltext{3}{National Astronomical Observatories, CAS, 20A Datun Road, Chaoyang District, Beijing 1000012, China.}
\altaffiltext{4}{National Institute for Astrophysics-OAPd, Vicolo dell'Osservatorio, 5, 35122, Padova, Italy; \\
Max-Planck Institut f\"ur Astrophysik, Karl-Schwarzschild-Str. 1, 85748 Garching, Germany.}
\altaffiltext{5}{Carnegie Institution of Washington, Las Campanas Observatory, Colina El Pino s/n, Casilla 601, Chile.}
\altaffiltext{6}{Department of Astronomy, University of California, Berkeley, CA 94720-3411, USA.}
\altaffiltext{7}{Carnegie Institution of Washington, 813 Santa Barbara Street, Pasadena, CA 91101, USA.}
\altaffiltext{8}{Department of Physics, Texas A\&M University, 4242 TAMU, College Station, TX 77843, USA.}

\begin{abstract}

\noindent
Optical and near-infrared photometry and optical spectroscopy are
reported for SN~2003bg, starting a few days after explosion and
extending for a period of more than 300 days.  Our early-time 
spectra reveal the presence of broad, high-velocity Balmer lines.
The nebular-phase spectra, on the other hand, show a remarkable
resemblance to those of Type Ib/c supernovae, without clear evidence
for hydrogen.  Near maximum brightness SN~2003bg displayed a
bolometric luminosity comparable to that of other Type I hypernovae
unrelated to gamma-ray bursts, implying a rather normal amount of
$^{56}$Ni production (0.1--0.2 \Msun) compared with other such
objects. The bolometric light curve of SN~2003bg, on the other hand,
is remarkably broad, thus suggesting a relatively large progenitor
mass at the moment of explosion. 
These observations, together with the large value of the kinetic energy of
expansion established in the accompanying paper \citep{mazzali09}, suggest 
that SN~2003bg can be regarded as a Type IIb hypernova.
\end{abstract}


\section{INTRODUCTION}

Stars with initial masses above $\sim 8$ \Msun\ are thought to end
their lives with the collapse of their core, when a compact remnant is
formed (a neutron star, or a black hole for the more massive stars) and
a powerful explosion expels the outer layers of the star. This
explosion is observed as a supernova (SN).

Depending on the degree of stripping of the outer layers before the
core collapses, supernovae (SNe) will have different characteristics
\citep[e.g.,][]{filippenko97}.  If a massive hydrogen envelope is
still present at the time of explosion, H lines dominate the spectrum
and H recombination dictates the light curve, and the SN is classified
as Type II. If the star loses its H-rich envelope but the He-rich
shell is still present, the SN is called a Type Ib event
\citep{elias85,porter87}, the light curve is not affected by the
recombination of hydrogen, and the spectrum shows strong He lines
attributed to non-thermal excitation caused by fast particles produced
in the \Nifs\ $\rightarrow$ \Cofs\ $\rightarrow$ \Fefs\ decay
\citep{lucy91}. If the He shell is also lost, the SN only shows lines
of heavier elements (O, Mg, Si, S, Ca, Fe) and is dubbed as a Type Ic
event \citep{wheeler86,wheeler87,filippenko90}. It is thought that the
loss of the outer H and He layers may be a consequence of a stellar
wind and/or binary interaction \citep{nomoto94a}.

The details of the explosion depend on the kinetic energy, the amount
of \Nifs~(i.e., the SN luminosity) produced in the explosion, and the
zero age main sequence (ZAMS) mass of the progenitor \citep[for SNe~II
  and SNe~Ib/c, respectively]{hamuy03a,nomoto05}.  Recent studies
suggest that stars more massive than $\sim 35$ \Msun\ lose both their
H and He shells prior to explosion, produce explosions with isotropic
energies exceeding 10$^{52}$ ergs \citep[for example]{mazzali07}, and
are often connected with gamma-ray bursts \citep[GRBs;][]{galama98}. A
signature of a high kinetic energy (more precisely, high kinetic energy
per unit mass) is a spectrum with broad lines \citep[for 
example]{iwamoto98}; in those cases where the total kinetic energy of
expansion is shown to indeed be very high, the object has been
dubbed a ``hypernova." Stars of somewhat smaller mass ($\sim
30$ \Msun) that still collapse to a black hole can also produce
broad-lined SNe~Ic but show no GRB [e.g., SN 1997ef; \citet{mazzali00}].
For H-rich SNe, on the other hand, very large kinetic energies have
not yet been observed. Only for SNe~Ib is there some evidence of unusually
high kinetic energies, albeit just from one event [SN 2008D; 
\citet{soderberg08,mazzali08}].

Here we present optical and near-infrared observations obtained of
SN~2003bg in the course of the ``Carnegie Type II Supernova Survey''
(CATS, hereafter). These data reveal that SN~2003bg is one of
the first broad-lined SNe~IIb ever observed. SNe~IIb are similar to SNe~Ib
but still show some traces of hydrogen in their early-time spectra,
which suggests the presence of a low-mass H layer prior to explosion;
the first known example was SN~1987K \citep{filippenko88}, and it too
had high expansion velocities. The prototypical object of this subclass 
is SN~1993J \citep{filippenko93,swartz93,filippenko94,nomoto94b}.

This paper is organized as follows. In Section 2 we describe the
observations. The analysis of the spectroscopic data and the
bolometric light curve are presented in Section 3.  Our conclusions are
summarized in Section 4. Detailed modeling of this supernova based on
our data is presented in the accompanying paper by \citet{mazzali09}, 
confirming the conclusion that SN 2003bg had very high kinetic energy
of expansion and thus can legitimately be classified as a hypernova.

\section{OBSERVATIONS}
\label{obs}

SN~2003bg was discovered on 2003 Feb. 25.7 (UT dates are used
throughout this paper) by R. Chassagne at Ste. Clotilde, Ile de
Reunion, with mag = 15 on an unfiltered CCD frame taken with a 0.30-m
automated telescope \citep{chassagne03}. The SN was located $16.3''$
west and $24.6''$ south of the nucleus of MCG-05-10-015, an SB(s)c
galaxy with a heliocentric recession velocity of 1367 km~s$^{-1}$
obtained from 21-cm neutral hydrogen measurements
\citep{theureau98}. According to Chassagne, nothing was visible at the
position of the SN on an unfiltered image taken on 2002 Nov. 7 with a
limiting magnitude of 18.0.

As soon as we learned of the discovery, we initiated detailed
follow-up observations of SN~2003bg at optical and near-infrared (NIR)
wavelengths, as part of the ``Carnegie Type II Supernova Survey.''
This program was carried out at Las Campanas Observatory during
2002--2003 with the main purpose to study nearby (redshift $z < 0.05$)
SNe~II. CATS used the 1-m Swope and 2.5-m du Pont telescopes to obtain
$UBVRIJ_sHK_s$ photometry and the Magellan Clay telescope for optical
spectroscopy. A total of 34 SNe~II were observed in 2002--2003 plus a
few SNe of other types. A detailed report of the 34 SNe~II is in 
preparation \citep{hamuy09}.  We also refer
the reader to \citet{hamuy06} describing the observational
procedures of the ``Carnegie Supernova Project'' (CSP), which are
nearly identical to those of CATS. In fact, CATS was a precursor to
the CSP initiated in 2004 with the aim to study SNe of all types. The
main difference between CATS and CSP is that the latter uses SDSS
$u'g'r'i'$ filters for optical imaging.

A complete journal of the photometric and spectroscopic observations
obtained for SN~2003bg is given in Table \ref{journal_tab}. In the
remainder of this section, we present the observation procedures, data
reductions, and the resulting data.

\subsection{Optical Photometry}

The optical imaging of SN~2003bg was obtained with $BVRI$
Johnson-Kron-Cousins filters using the Swope 1-m $f/7$ telescope
facility Direct CCD Camera, the Wide Field Re-imaging CCD Camera
(WFCCD) on the 2.5-m $f/7.5$ du Pont telescope, and the Low-Dispersion
Survey Spectrograph \citep[LDSS-2;][]{allington94} on the 6.5-m $f/11$
Magellan Clay telescope, all at Las Campanas Observatory (LCO).  A few
additional optical images were obtained with the 0.9-m telescope at
Cerro Tololo Inter-American Observatory (CTIO). The SN observations
started on 2003 Mar. 2 and extended through 2004 Jan. 15, covering 320
days of the SN evolution.  For galaxy subtraction, we used images
obtained with the Direct CCD Camera on the 2.5-m du Pont telescope on
2004 Nov. 11 and 2005 Feb. 13 when the SN was no longer visible. This
instrument delivers images with a typical image quality of $0.7''$
full width at half-maximum intensity (FWHM), thus providing galaxy 
templates with better image quality than any of the SN+galaxy images.

All of the photometric reductions were performed with customized
IRAF\footnote{IRAF is distributed by the National Optical Astronomy
Observatories, which are operated by the Association of Universities
for Research in Astronomy, Inc., under cooperative agreement with
the National Science Foundation.} 
scripts. The reductions started
with the subtraction of the galaxy templates, which involved (1)
determining the coordinate transformation between the two images and
registering the template to the SN+galaxy image, (2) degrading the
point-spread function (PSF) of the template to that of the SN+galaxy
image, (3) matching the flux scale of the template to that of the
SN+galaxy image, (4) subtracting the modified template from the
SN+galaxy image, and (5) extracting a small box around the SN from the
subtracted image and inserting it into the original SN+galaxy image. As
a result of this procedure the SN ended up lying on a smooth
background, thus allowing us to reliably measure the SN flux.

\begin{figure}
\plotone{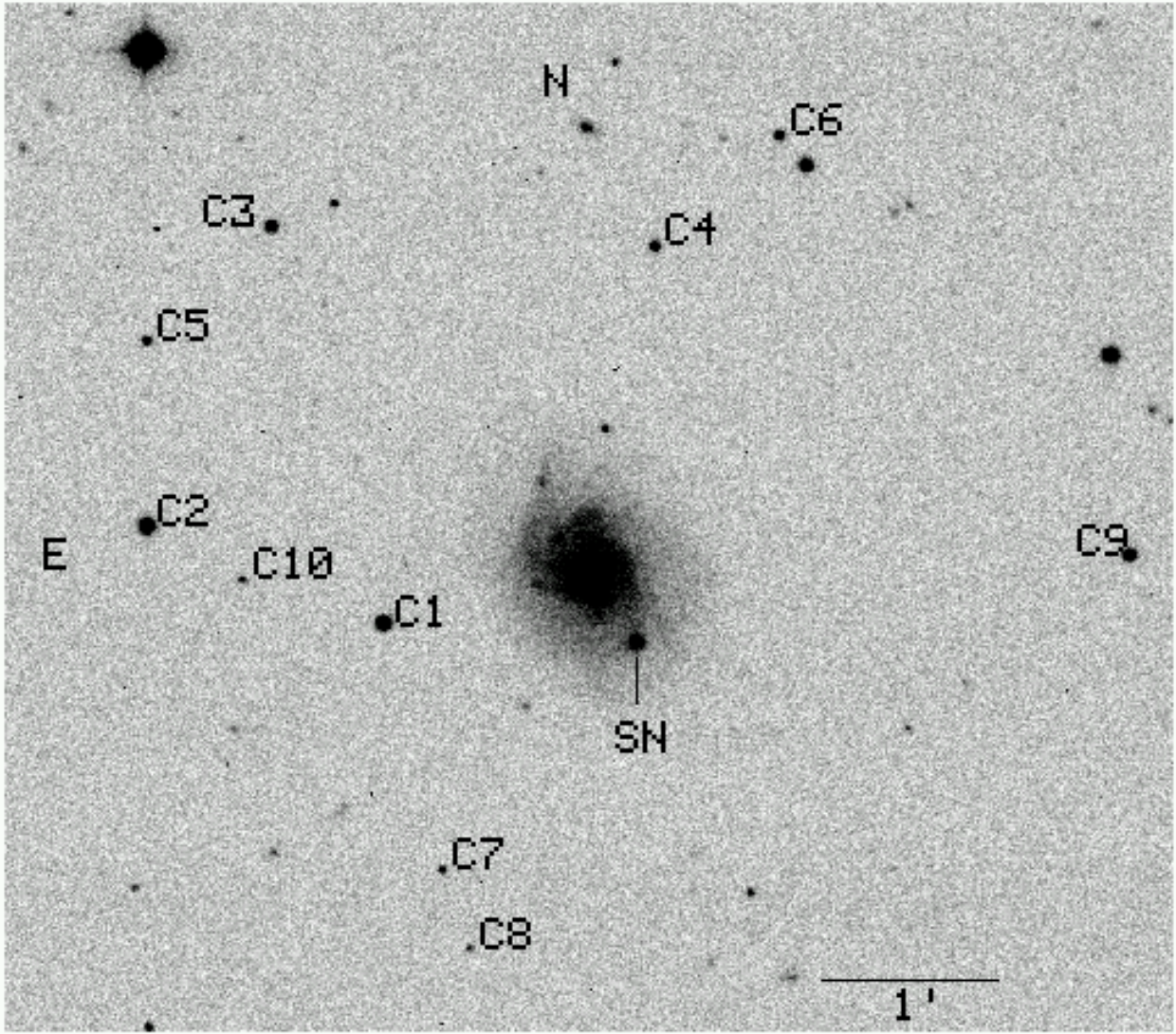}
\caption{The field of SN~2003bg observed on 2003 Jul. 25 with the
  Swope 1-m telescope and a $V$ filter. North is up and east is to the
  left. The supernova is marked to the SW of the host galaxy. Ten
  comparison stars used to derive differential photometry of the SN
  are labeled.  The scale is shown with an horizontal line near the
  bottom.
}
\label{fchart_fig}
\end{figure}

The next step consisted of establishing a photometric sequence of
local standard stars in the field of SN~2003bg. For this purpose we
employed observations with the Swope telescope of \citet{landolt92}
standard stars during eight photometric nights. Figure
\ref{fchart_fig} shows the SN field and the ten selected stars, and
Table \ref{sequence_tab} lists the average $BV(RI)_{KC}$ magnitudes
for these stars.

SN magnitudes in the standard Johnson-Kron-Cousins system were
obtained differentially relative to the comparison stars using PSF
photometry on the galaxy-subtracted images. For this purpose, we
employed all the stars of the photometric sequence to determine an
average PSF for every CCD image, and we fitted the resulting PSF to
the SN and the standards to a radius of $2''$.  The sky was estimated
locally for each star from an annulus centered around the star with an
inner radius of $7''$ and a width of $2''$.  The instrumental
magnitudes of the SN were converted to the standard system using the
following equations:

\begin{equation}
B~=~b~+~ct_b~(b~-~v)~+~zp_b,
\label{B_eq}
\end{equation}
\begin{equation}
V~=~v~+~ct_v~(v~-~i)~+~zp_v,
\label{V_eq}
\end{equation}
\begin{equation}
R~=~r+~ct_r~(v~-~r)~+~zp_r,~{\rm and}
\label{R_eq}
\end{equation}
\begin{equation}
I~=~i+~ct_i~(v~-~i)~+~zp_i.
\label{I_eq}
\end{equation}

\noindent In these equations $BVRI$ (left-hand side) are the published
magnitudes in the standard system \citep{landolt92}, $bvri$
(right-hand side) correspond to the natural-system magnitudes,
$ct_i$ is the color term, and $zp_i$ is the zeropoint for filter 
$i$.\footnote{In the case of the LDSS-2 instrument which does not have 
an $I$-band filter, we used $(b-v)$ instead of $(v-i)$ in equation 
\ref{V_eq}.} For all four optical cameras we employed average color 
terms determined on multiple nights (listed in Table \ref{coefficients_tab}), 
solving only for the photometric zeropoints.

We emphasize that our photometric reductions did not include
shutter-time corrections because at the time of CATS we did not have
an assessment of this effect.  While the differential photometry
technique is immune to the shutter-time error, the calibration of the
local photometric sequence performed with the Swope telescope could be
potentially affected by this effect since the exposures of the
\citet{landolt92} standards are much shorter than those of the SN
field. During the CSP survey in 2004 we measured this effect and
determined that for the curtain shutter design of the CCD camera on
the Swope telescope, the error introduced in the photometry is a
single constant error across the field amounting to 0.08~s
\citep{hamuy06}. For our shortest exposures of the Landolt fields
(3~s), the neglect of this correction introduces an overestimate of
2.6\% in the observed fluxes. The final error in the calibration of
the local standards is certainly much lower than this owing to the
fact that the nightly photometric calibrations were derived from
$\sim$20 Landolt stars, most of which observed with longer
exposures. We thus conclude that neglecting the shutter correction
makes the SN appear systematically fainter, but the effect is
$\lesssim$ 0.01 mag.

\begin{figure}
\plotone{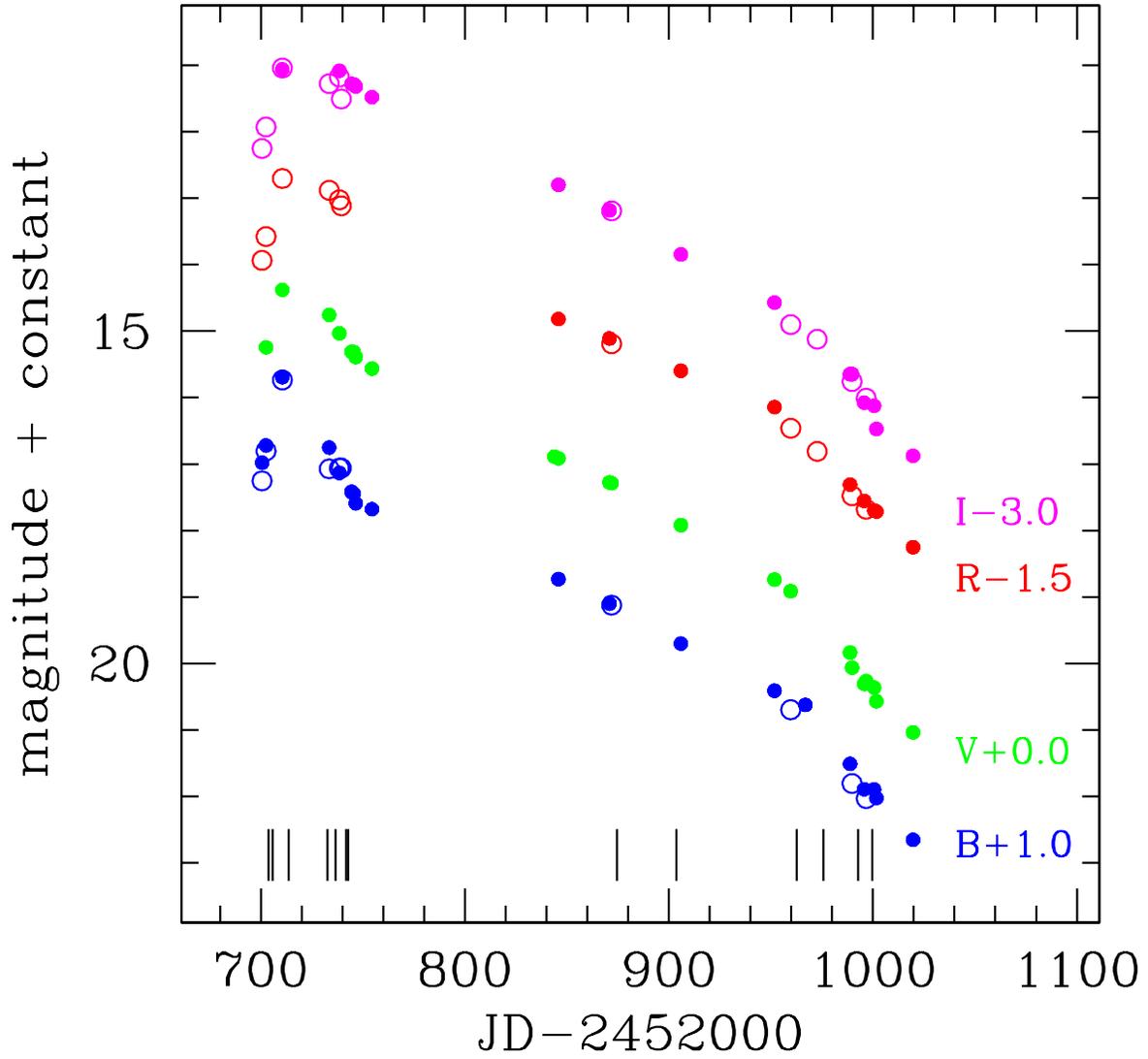}
\caption{$BV(RI)_{KC}$ light curves of SN~2003bg. Open circles
  indicate synthetic magnitudes computed from the spectra, while
  filled circles indicate magnitudes computed from the direct images.
  The vertical bars indicate the epochs of our optical spectra.  
}
\label{photometry_fig}
\end{figure}

We also note that we did not include CCD linearity corrections for any
of our instruments since we lacked such information during CATS.
After the completion of the project we did carry out such measurements
with the Swope ``Site\#3'' CCD in the course of the CSP, and we
reached the conclusion that the response of the detector departed 6\%
from linearity at 23,000 ADU~pixel$^{-1}$ (57,500 e$^-$~pixel$^{-1}$)
\citep{hamuy06}. Based on that information we have empirically
determined that the effect of neglecting this correction introduces a
systematic error in the instrumental magnitudes of $\sim$0.005 mag per
magnitude unit.  We therefore expect a differential bias in the SN
flux with respect to the local standards.  The sign and net error in
the SN magnitude will depend on whether the SN is brighter or fainter
than the average flux of the local standards. During our follow-up
observations, the SN varied in the range $V = 14.4$--21.0 mag (see
Figure \ref{photometry_fig}), while the average local standard had $V
= 17.8$ mag (see Table \ref{sequence_tab}), so we expect to
overestimate the SN flux by $V = 0.017$ mag around maximum light and
underestimate the SN flux by $V = 0.016$ mag toward the end of the
campaign. These corrections are on the same order as the precision of
the final photometry, and so we have not applied them.


The resulting $BV(RI)_{KC}$ magnitudes of SN~2003bg are listed in
Table \ref{photometry_tab}.  The final uncertainty in the SN magnitude
was the instrumental error in the PSF fit (assuming a minimum of 0.015
mag). We neglected errors due to the transformation to the standard
system since the uncertainties in the color term and zeropoint
were well below 0.015 mag. This table also includes synthetic $BRI$
magnitudes computed from our spectra (see below for details) in which
case we adopted a conservative uncertainty of 0.1 mag.  

Figure \ref{photometry_fig} shows the $BVRI$ light curves, all of
which reveal that the SN was caught during an early phase of steep
brightening. We estimate that the SN reached a peak of $V_{\rm max} =
14.2$ mag at JD 2,452,716--720 (2003 Mar 17--21). This relatively large 
uncertainty in
time is caused by the fact that the light curve around maximum was not
sampled in great detail.  Peak brightness was followed by a steep
decline of $\sim$1.3 mag in 30 days. Over the next 150 days the SN
showed a slow but steady decline at 0.015 mag day$^{-1}$. Starting
around JD 2,452,970 ($\sim$252 days past maximum), there was an evident
increase in the decline rate which might be due to the formation of
dust in the SN ejecta, as was the case for SN~1987A $\sim$500 days
after explosion \citep{suntzeff90}.

\begin{figure}
\plotone{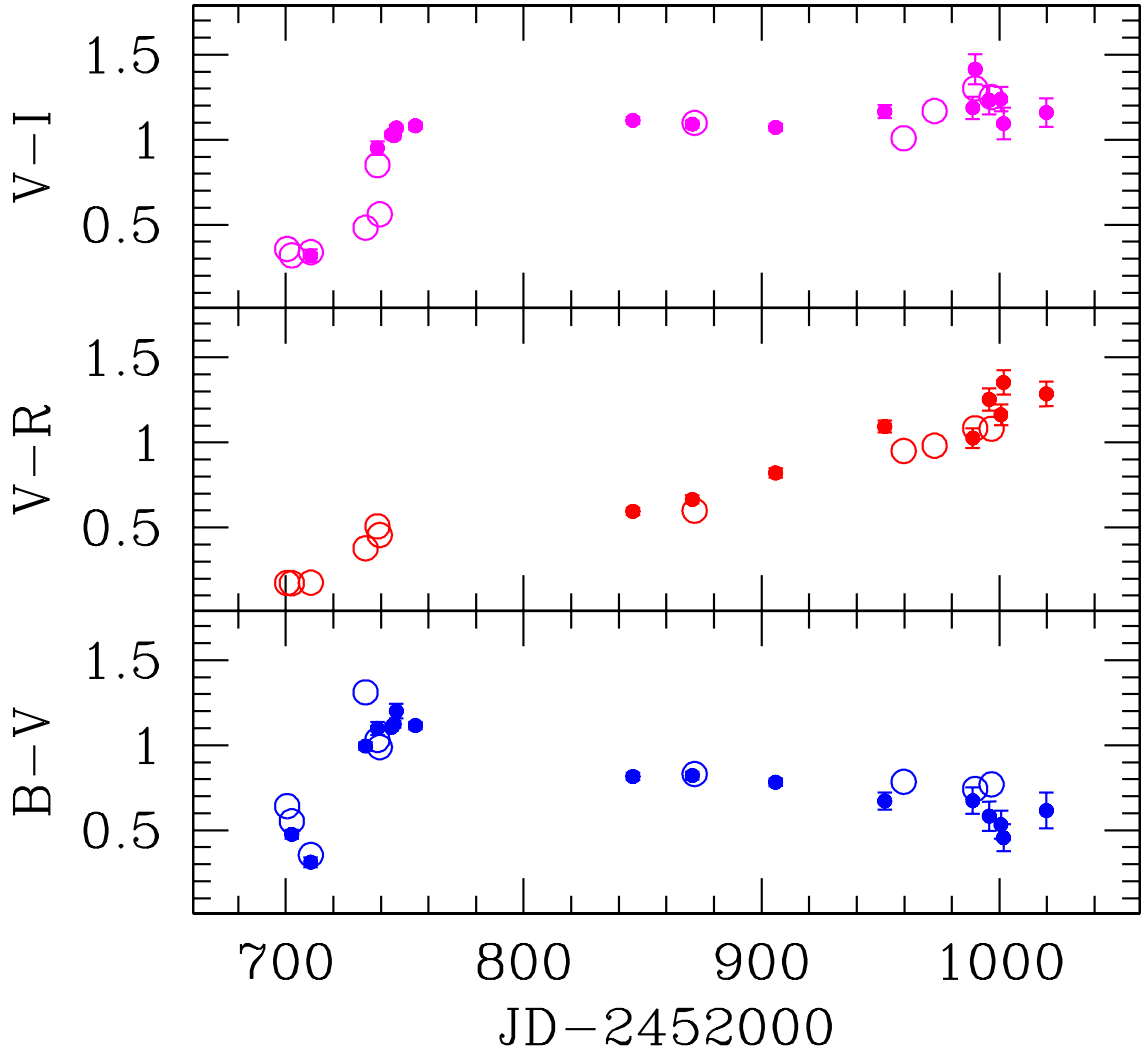}
\caption{$B-V$, $V-R$, and $V-I$ color curves (mag) of SN~2003bg. Open
  circles indicate synthetic colors computed from the spectra, while
  filled circles indicate colors computed from the observed
  photometry.  
}
\label{color_fig}
\end{figure}

The color curves are shown in Figure \ref{color_fig}. Prior to maximum
the SN became gradually bluer in $B-V$, after which it grew redder in
all three colors through JD 2,452,740. After that the SN became bluer
in $B-V$ and redder in $V-I$.

\subsection{Infrared Photometry}

We obtained $J_sHK_s$ images of SN~2003bg with the Swope telescope IR
Camera (C40IRC) and the Wide Field IR Camera (WIRC) on the du Pont
telescope \citep{persson02}.  C40IRC is equipped with a single $256
\times 256$ pixel Rockwell NICMOS-3 HgCdTe array with a $0.6''$
pixel$^{-1}$ scale, which corresponds to a field of view (FOV) of
$2.5' \times 2.5'$. WIRC is equipped with four $1024 \times 1024$
pixel Rockwell HAWAII-1 HgCdTe arrays forming a $2 \times 2$ square
footprint with a $175''$ center-to-center spacing in the reimaged
telescope focal plane.  Each array covers a FOV of $\sim3.3' \times
3.3'$ with a $0.196''$ pixel$^{-1}$ scale.

Sky flats were taken with C40IRC during evening twilight. Dark frames
were obtained just after closing the dome at the end of the
night. With WIRC we instead obtained dome flats with the dome closed
at the end of the night followed by dark frames.  With both
instruments we observed the SN by taking sequences of dithered
exposures ranging between 20~s and 90~s depending on the brightness of
the SN.  With C40IRC we observed SN~2003bg along with star c1 (Figure
\ref{fchart_fig}) in order to conduct differential photometry. The
larger FOV of WIRC allowed us to include star c1 and a few other field
stars. During six clear WIRC nights, we observed one to four standard
stars from \citet{persson98} in order to calibrate star c1.

The C40IRC and WIRC reductions consisted of subtracting the dark
frames from the flat images and dividing the normalized flats into all
the science frames.
Every time standard stars were observed with WIRC, we measured
instrumental magnitudes through a standard aperture of $5''$ radius
\citep{persson98}, with a sky annulus $5''$ to $7''$ from the star. We
corrected such magnitudes for atmospheric extinction using the
canonical values given by \citet{persson98}: $k_J=0.1$, $k_H=0.05$,
and $k_K=0.08$. A photometric zeropoint was then derived for each
filter, assuming zero color terms since the instrument detector and
filters used with WIRC are essentially the same as in
\citet{persson98}.  The photometric transformations derived from these
observations were then applied to star c1. From the six independent
measurements we obtained the following average magnitudes: $J_s =
14.948 \pm 0.018$, $H = 14.610 \pm 0.023$, and $K_s = 14.594 \pm
0.020$. These values are in excellent agreement with the values $J_s =
14.928 \pm 0.047$, $H = 14.604 \pm 0.052$, and $K_s = 14.535 \pm
0.096$ in the 2MASS All-Sky Catalog of Point Sources
\citep{skrutskie06}.

We then performed differential photometry of the SN relative to star
c1. To improve the instrumental precision, we measured aperture
photometry of both stars using a small aperture that varied between
$1''$ and $1.5''$ depending on the seeing.  The instrumental
magnitudes were converted to the standard system using the zeropoints
derived for each filter from star c1. The resulting $J_sHK_s$
magnitudes of SN~2003bg are listed in Table
\ref{ir_photometry_tab}. The final uncertainty in the SN magnitude is
the sum in quadrature of the statistical errors in the instrumental
magnitudes of the SN and star c1, and the errors in the standard
magnitudes of star c1.  At this point we have ignored possible errors
caused by host-galaxy contamination in the SN aperture because we lack
galaxy templates in the IR filters.  Given the smooth background on
which the SN lies, we believe this error is negligible.

\begin{figure}
\plotone{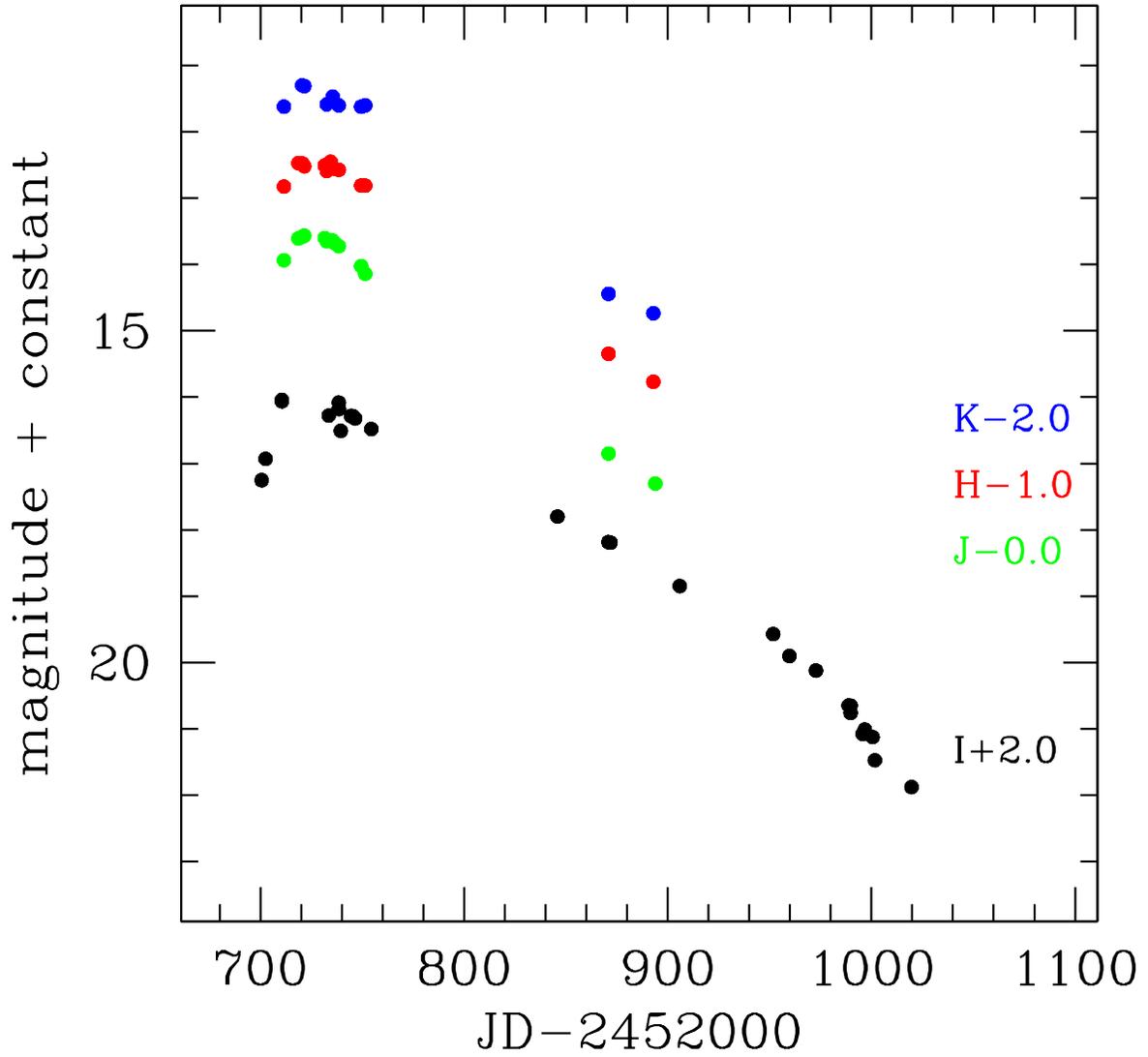}
\caption{$J_sHK_s$ light curves of SN~2003bg. The $I$-band light curve is 
also shown to guide the eye.
}
\label{IRphotometry_fig}
\end{figure}

Figure \ref{IRphotometry_fig} shows the resulting IR light curves of
SN~2003bg. At these wavelengths the epoch around maximum brightness
was very well sampled.  The SN was found $\sim$15 days before maximum,
which occurred at JD 2,452,726 $\pm$ 1 with $J_{\rm max} = 13.54$,
$H_{\rm max} = 13.45$, and $K_{\rm max} = 13.32$ mag. The shapes of the IR
light curves are similar to those at optical wavelengths.

\begin{figure}
\plotone{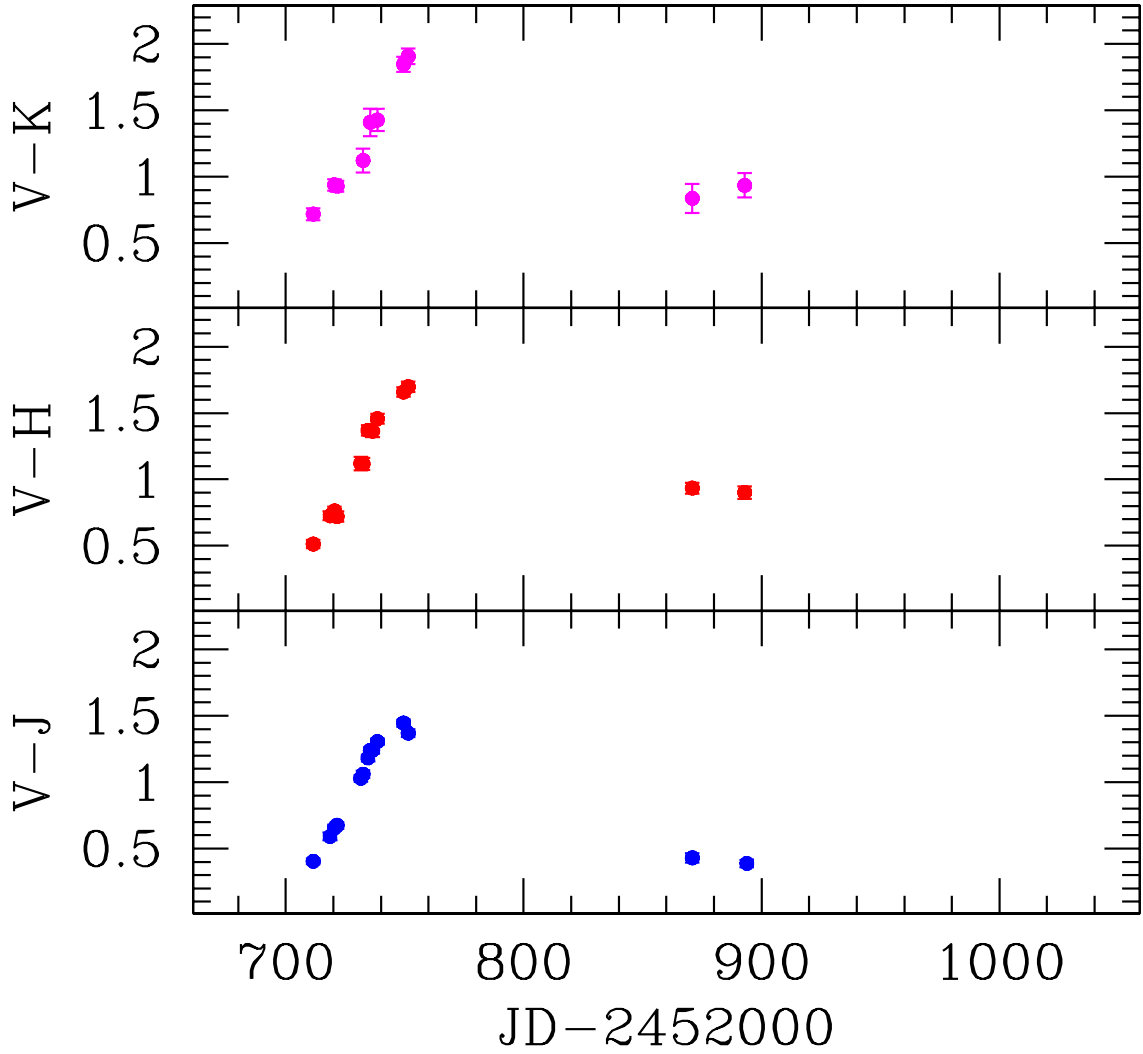}
\caption{$V-J$, $V-H$, and $V-K$ color curves (mag) of
  SN~2003bg. Since the optical and IR magnitudes were obtained at
  different times, it was necessary to interpolate the $V$ magnitudes
  shown in Figure \ref{photometry_fig} to the time of the IR
  observations.
}
\label{color_IR_fig}
\end{figure}

In Figure \ref{color_IR_fig} we present the $V$--IR colors. In all
three colors, the SN evolved from blue to red through JD 2,452,740 and
then became systematically bluer through the last observations on
JD 2,452,900.

\subsection{Spectroscopy}

A total of 14 optical spectra were obtained of SN~2003bg. Table
\ref{optspec_tab} gives a journal of the spectroscopic observations.
The majority of our spectra were obtained with the 2.5-m du Pont
telescope using the WFCCD instrument in its spectroscopic long-slit
mode.  A 400 line mm$^{-1}$ blue grism was employed with the
Tektronix $2048 \times 2048$ pixel CCD to provide a wavelength
coverage of 3800--9330~\AA\ at a dispersion of
3.0~\AA~pixel$^{-1}$. With the $\sim1.6''$ slit width used for the SN
observations, this setup gave a 
FWHM resolution of $\sim$6.0~\AA.  On the 2.5-m telescope we also
obtained spectra with the Las Campanas Modular Spectrograph.  This
instrument uses a SITe $1752 \times 572$ pixel CCD with 15 $\mu$m
pixel$^{-1}$ and a 300 line mm$^{-1}$ grating (blazed at 5000~\AA).
The resulting spectral coverage is $\sim$3780--7280~\AA\ at a
dispersion of 2.45~\AA~pixel$^{-1}$.  A slit width of $1''$ was used
for the SN observations, which gave a FWHM resolution of $\sim$7~\AA.

On two occasions we used the 6.5-m Magellan Clay telescope with
LDSS-2.  For these observations, a 300 line mm$^{-1}$ grism blazed at
5000~\AA\ was employed, providing wavelength coverage of
3600--9000~\AA\ at a dispersion of 5.3 \AA~pixel$^{-1}$.  With a $1''$
slit, this translates to a FWHM resolution of $\sim13.5$~\AA.  We also
obtained a single spectrum using the Low Resolution Imaging
Spectrometer \citep[LRIS;][]{oke95} on the Keck-I 10-m telescope.  The
D560 dichroic beamsplitter was used to split the spectrum near
$\sim5500$~\AA.  The $1''$ slit combined with the 400 line mm$^{-1}$
grism blazed at 3400~\AA\ on the blue side of the spectrograph and the
400 line mm$^{-1}$ grating on the red side to give resolutions of 7
and 6~\AA, respectively.

We emphasize that, with the exception of the LRIS spectrum, for none
of these spectra did we employ a filter to block second-order light,
which means that there could be second-order contamination redward of
$\sim$6800~\AA. With the WFCCD and LDSS-2 we observed the SN with the
slit aligned along the parallactic angle \citep{filippenko82} to
reduce the effects of atmospheric dispersion.  However, owing to a
misinterpretation of the observing instructions, with the Modular
Spectrograph we oriented the slit perpendicular to this angle,
introducing significant errors in the relative spectrophotometry.  The
Keck/LRIS spectrum was obtained at a high airmass (4.2) with mediocre
and variable seeing conditions, and SN~2003bg was partially vignetted
by the dome shutter, so the absolute flux scale is highly uncertain.
However, the slit was aligned along the parallactic angle, so the
relative spectral shape is reliable.

\begin{figure}
\plotone{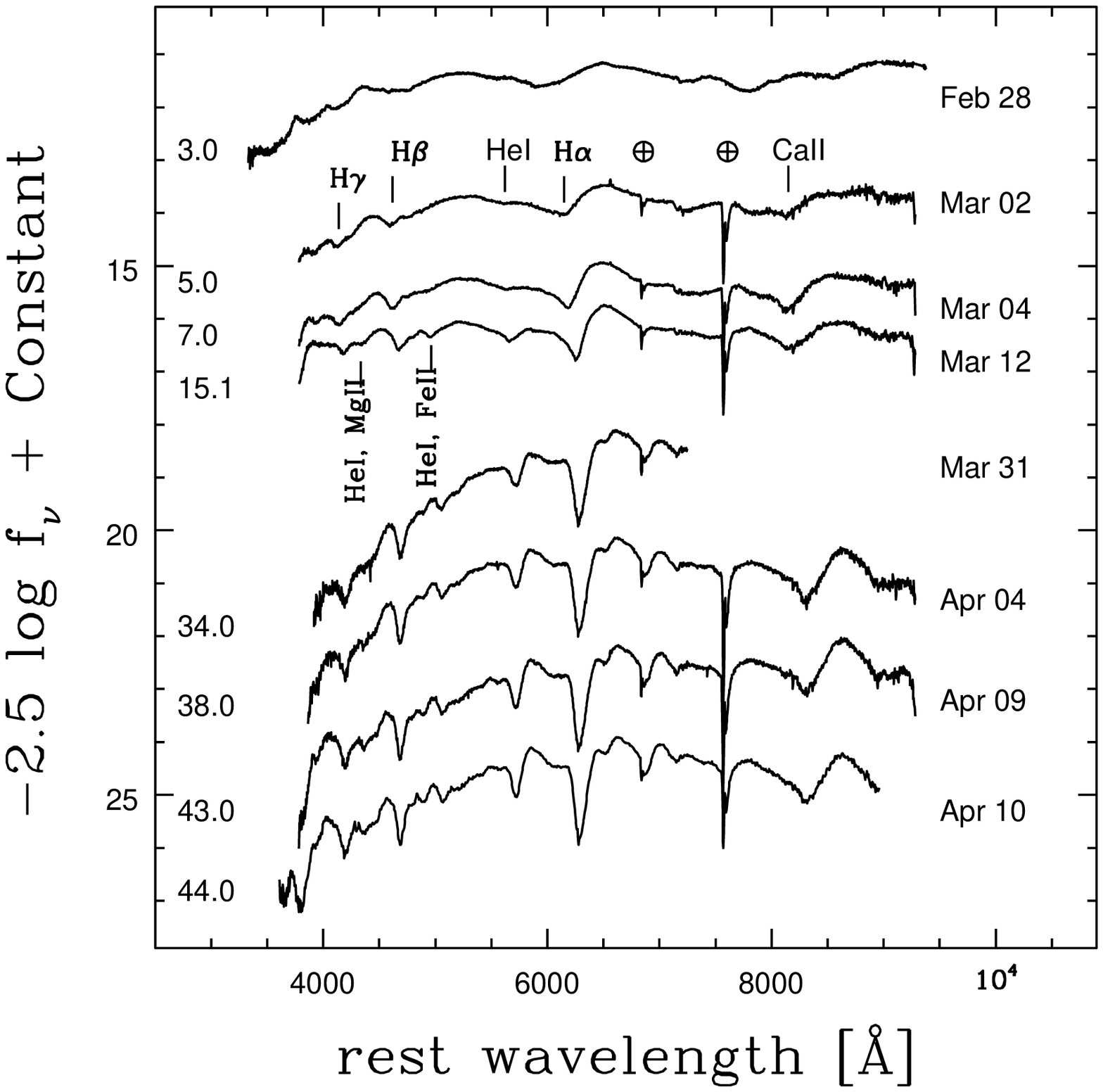}
\caption{Spectroscopic evolution of SN~2003bg during the photospheric
  phase.  The wavelengths of the spectra were shifted to the SN rest
  frame using a heliocentric recession velocity of 1367 km~s$^{-1}$.
  The labels to the left of the spectra indicate the rest-frame days
  elapsed since explosion (assumed to be on JD 2,452,695.5). Telluric
  features are indicated with the $\oplus$ symbol.  The spectrum
  obtained on Mar. 31 was obtained with the slit oriented perpendicular
  to the parallactic angle, which introduced significant errors in the
  relative spectrophotometry.
}
\label{spec1a_fig}
\end{figure}

\begin{figure}
\plotone{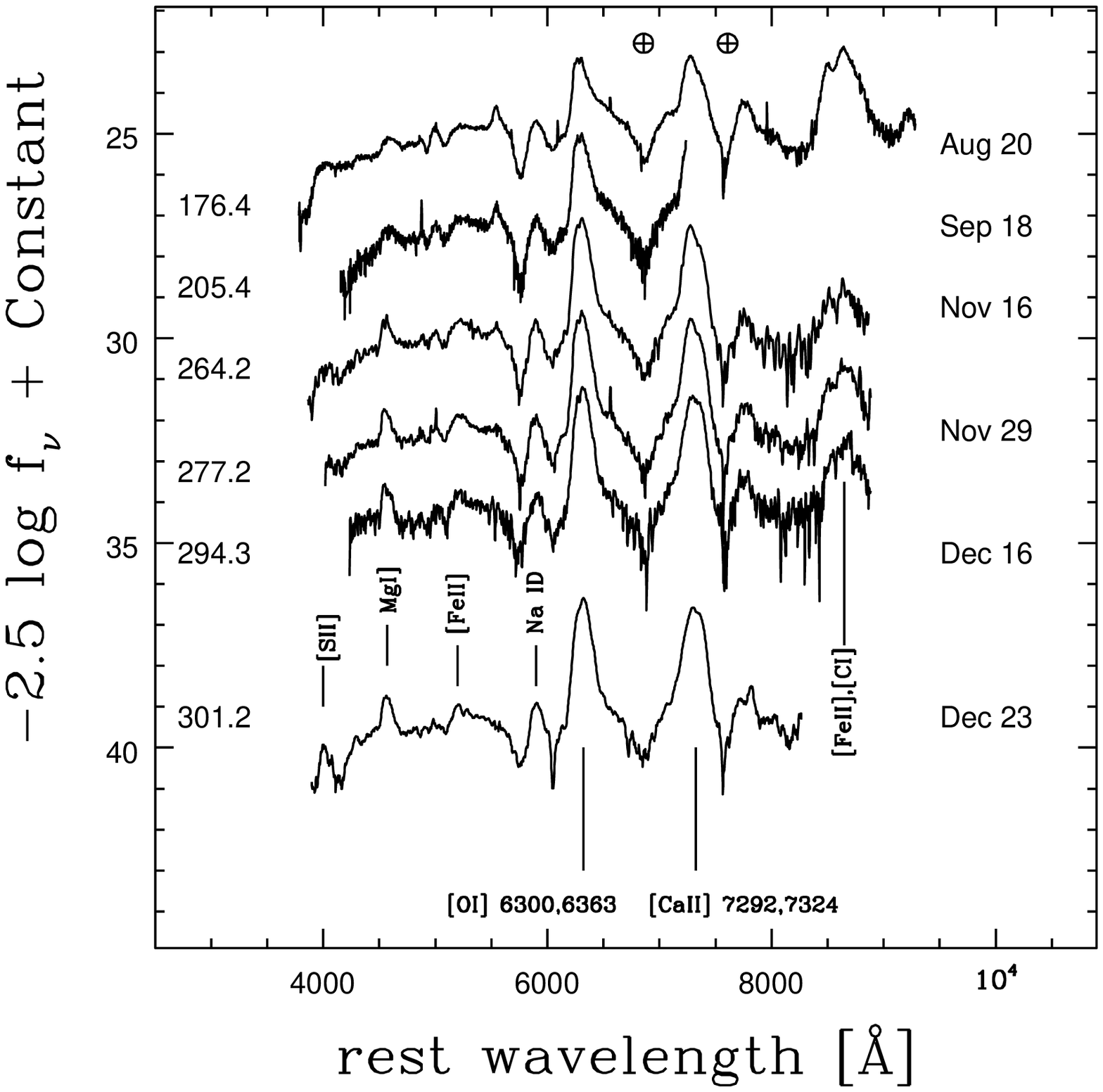}
\caption{Spectroscopic evolution of SN~2003bg during the nebular
  phase.  The wavelengths of the spectra were shifted to the SN rest
  frame using a heliocentric recession velocity of 1367 km~s$^{-1}$.
  The labels to the left of the spectra indicate the rest-frame days
  elapsed since explosion (assumed to be on JD 2,452,695.5). Telluric
  features are indicated with the $\oplus$ symbol.  The spectrum on
  Sep. 18 was obtained with the slit oriented perpendicular to the
  parallactic angle, which introduced significant errors in the
  relative spectrophotometry.
}
\label{spec1b_fig}
\end{figure}

All of the spectra were wavelength and flux calibrated using
comparison lamp and standard-star observations \citep{hamuy94b}. No
attempt was made to remove telluric absorption lines, except in the
Keck/LRIS spectrum.  Figures \ref{spec1a_fig} and \ref{spec1b_fig} show
the series of spectra of SN~2003bg obtained during the photospheric
and nebular phase, respectively.  The wavelengths of the spectra were
shifted to the SN rest frame using a heliocentric recession velocity
of 1367 km~s$^{-1}$ measured for the host galaxy \citep{theureau98}.
The labels to the left of the spectra indicate the rest-frame days
elapsed since explosion [assumed to be on JD 2,452,695.5 (Feb 25.0)
  based on the spectrum fitting by \citet{mazzali09}].  Telluric
features are indicated with the $\oplus$ symbol.  It is evident in
these figures that the shape of the two spectra obtained with the
Modular Spectrograph on Mar. 31 and Sep. 18 differs from that of the
remaining spectra owing to the wrong orientation of the slit with
respect to the parallactic angle.

We checked the spectrophotometric quality of our spectra by convolving
them with the $BVRI$ filter passbands described by \citet{bessell90}
and computing synthetic magnitudes.  Figure \ref{color_fig} compares
the $B-V$, $V-R$, and $V-I$ synthetic colors (open circles) with the
colors computed from the observed magnitudes. In general this
comparison reveals very good agreement (within 0.1 mag) between the
observed and synthetic colors. The most evident discrepancies occur in
$V-I$ (near JD 2,452,740). This is not surprising considering that we
did not use an order-blocking filter in our observations and, therefore, our
spectra must suffer from second-order contamination redward of 6800
\AA. We note that we excluded from this comparison the two spectra
obtained with the Modular Spectrograph which were not obtained along
the parallactic angle.

Having checked the relative spectrophotometry of our spectra, we
proceeded to combine the synthetic colors with our observed $V$
magnitudes (interpolated to the epoch of our 
spectroscopy)\footnote{For the first spectrum, which was obtained 
two days prior to the first photometric point, it proved necessary to
extrapolate the $V$-band light curve using a third-order polynomial
fit to the data around maximum light} in order to derive $BRI$
synthetic magnitudes and complement the observed light curves. The
resulting synthetic magnitudes are shown with open circles in Figure
\ref{photometry_fig}. This plot reveals that the spectrophotometry is
a very useful complement to the observed light curves, especially
before maximum light.

\section{ANALYSIS}
\label{res}

\subsection{Spectroscopic Analysis During the Photospheric Phase}

The spectroscopic evolution of SN~2003bg is analyzed in detail in the
companion paper by \citet{mazzali09}. Here we present only a brief
description of the main lines detected and the deduced velocities.

Our first spectrum, obtained on Feb. 28 with Keck/LRIS (see Figure
\ref{spec1a_fig}), is characterized by a blue continuum and several
broad P~Cygni profiles. \citet{filippenko03} claimed that SN~2003bg
was very similar to Type Ic hypernovae SN 1997ef, SN 1998bw, SN
2002ap, and SN 2002bl, perhaps two weeks after explosion; we see here
that it was actually probably closer to just 3 days past
explosion. Although one might be tempted to associate the absorption
trough near 6000~\AA\ with H$\alpha$ having an expansion velocity of
$\sim$25,000 km~s$^{-1}$, the model of \citet{mazzali09} shows that H
lines make only a minor contribution at this time; instead, the
6000~\AA\ trough is dominated by Si~II $\lambda$6355. Similarly,
\citet{mazzali09} find that He lines are not easily visible at this
time. In addition to Si~II, there are features produced by Fe~II,
Co~II, O~I, Ca~II (the near-IR triplet), and Mg~II. Thus, it cannot be
concluded that SN 2003bg was a SN~IIb based on this spectrum alone.

By Mar. 2, H$\alpha$ has appeared and is now comparable to (and blended
with) Si~II $\lambda$6355 \citep{mazzali09}. The H$\alpha$ expansion
velocity, measured from the absorption minimum, is 19,300 km s$^{-1}$;
however, the blend with Si~II $\lambda$6355 artificially increases this
over its true value. The weak feature at 5620 \AA\ is probably
He~I $\lambda$5876.  A strong feature due to the Ca~II near-IR triplet 
is observed around 7800--8300~\AA, as well as a feature at 4780~\AA\ 
which is attributed to Fe~II $\lambda\lambda$5018, 5169.

The spectrum obtained on Mar. 4 is generally similar to the previous
one. It reveals the presence of strong, broad H$\alpha$ emission with
a P~Cygni absorption component having an expansion velocity of 17,700
km~s$^{-1}$ at its minimum (though still probably an overestimate due
to residual blending with Si~II $\lambda$6355, which is now weaker
than H$\alpha$). A P~Cygni absorption line corresponding to H$\beta$
has developed at an expansion velocity of 14,800
km~s$^{-1}$. \citet{hamuy03} noted that the overall appearance of the
spectrum of SN~2003bg was similar to that of other broad-lined SNe~Ic,
but the presence of strong H features distinguished it, making it the
first broad-lined SN~II. [However, as noted below, the SN~IIb 1987K
\citep{filippenko88} actually had a comparable expansion velocity at
the same phase.]

On Mar. 12 (the spectrum nearest the time of maximum brightness), a
prominent He~I $\lambda$5876 line is present with an expansion velocity of 
10,900 km~s$^{-1}$, while the H$\alpha$ velocity has dropped to 14,200 
km~s$^{-1}$, close to (or perhaps even somewhat less than) that of
SN~IIb 1987K near maximum light \citep{filippenko88}. A prominent 
absorption at rest wavelength 4962~\AA\ has emerged, possibly due 
to a blend of He~I $\lambda$5048, Fe~II $\lambda$5018, and Fe~II 
$\lambda$5169. The feature at 4337~\AA\ has been identified by 
\citet{mazzali09} as a blend of He~I $\lambda$4471 and Mg~II 
$\lambda$4481. The Ca~II near-IR triplet remains strong.

By Apr. 4 the continuum has become much redder. The Balmer lines are
still present with lower expansion velocities (12,900 km~s$^{-1}$ for
H$\alpha$). He~I $\lambda$5876 is much more conspicuous at this epoch
with a velocity of 8000 km~s$^{-1}$. The Ca~II near-IR triplet has
developed an evident P~Cygni profile. The spectrum obtained on Apr. 10
is the last of the photospheric phase and is not very different than
the previous one.

\begin{figure}
\plotone{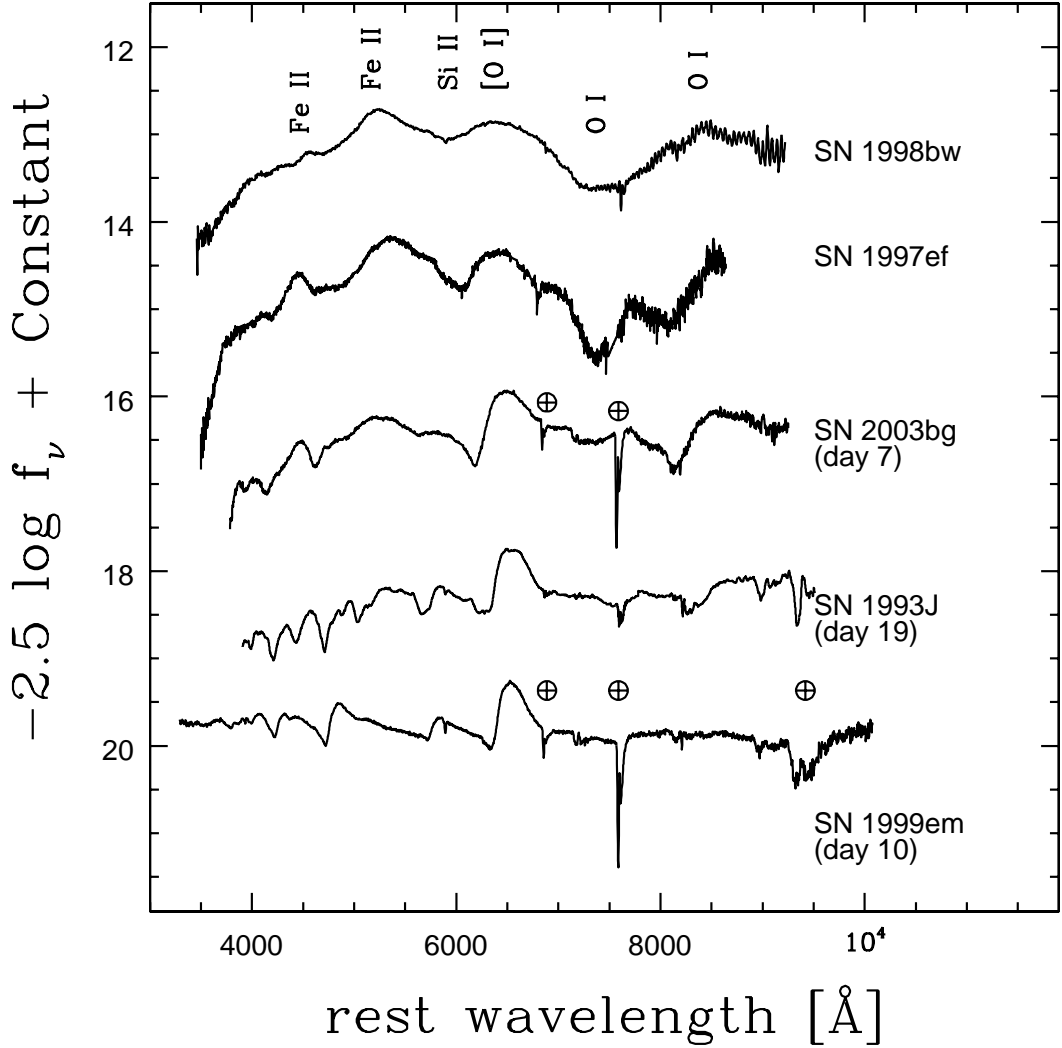}
\caption{Comparison of maximum-light spectra of hypernova SN 1998bw
  \citep{patat01}, hypernova SN 1997ef
  \citep{garnavich97a,garnavich97b}, SN 2003bg, Type IIb SN~1993J
  \citep{filippenko93}, and Type II plateau SN 1999em \citep{hamuy01}.
  Telluric features are indicated with the $\oplus$ symbol.
}
\label{spec4_fig}
\end{figure}

Figure \ref{spec4_fig} compares the maximum-light spectrum of
SN~2003bg with contemporaneous spectra of other core-collapse SNe
(II-P, IIb, Ic unrelated to GRBs, and GRB-related Ic).  Near maximum
brightness, SN~2003bg stands out for its broad features compared to
the typical Type II-P SN~1999em [\citet{hamuy01,leonard02,elmhamdi03}] and 
the Type IIb SN~1993J \citep{filippenko93}.  As noted by \citet{filippenko03}, 
the spectrum of SN~2003bg is somewhat reminiscent of that shown by the
broad-lined SNe~Ib/c. This can be seen in Figure \ref{spec4_fig} from
the comparison with the maximum-light spectra of SN~1997ef \citep[][a
broad-lined supernova not associated with a GRB]{garnavich97a,garnavich97b} 
and the GRB-connected SN~1998bw \citep{patat01}.

\begin{figure}
\plotone{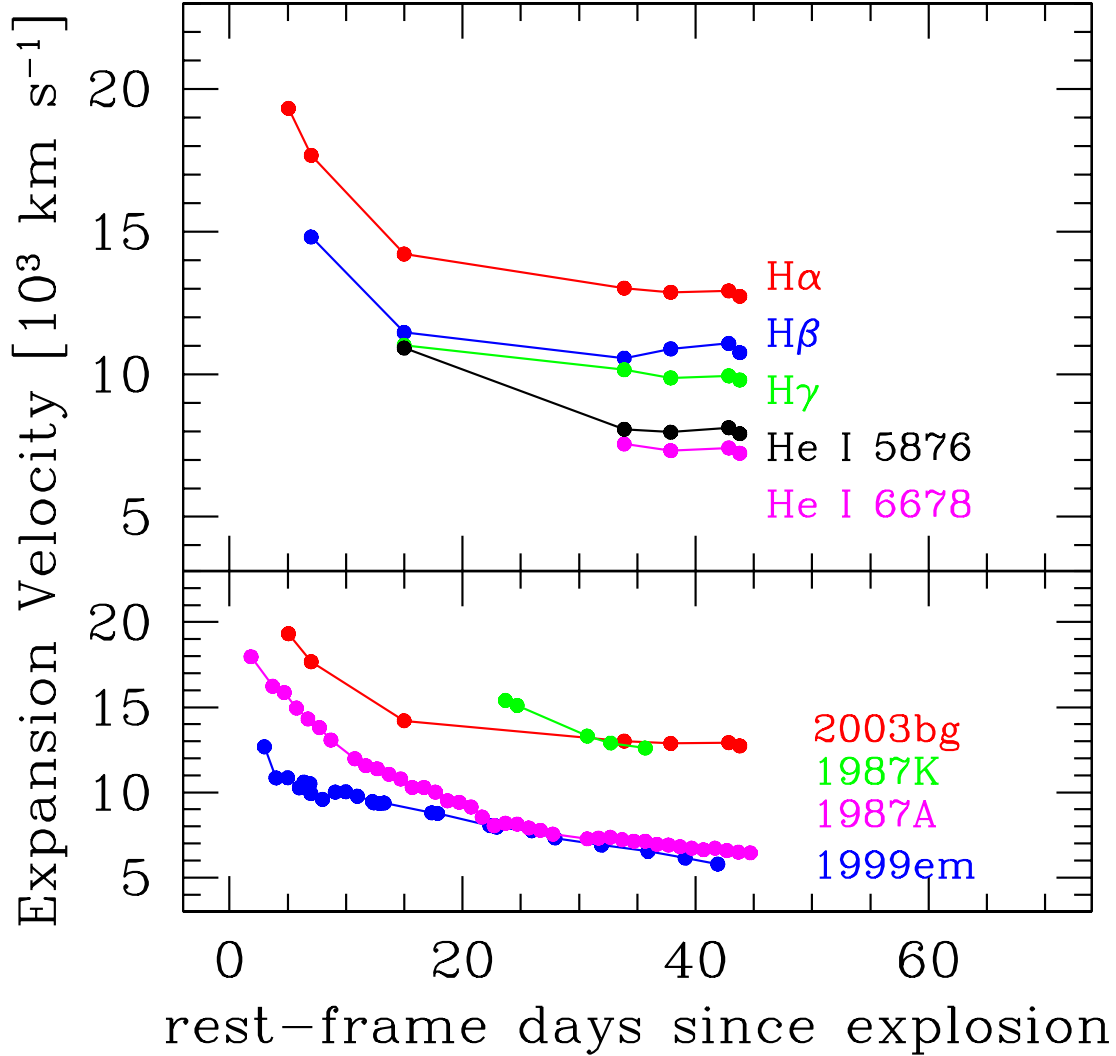}
\caption{(top) Expansion velocities for SN~2003bg measured from the
minimum of the H$\alpha$, H$\beta$, H$\gamma$, He~I $\lambda$5876,
and He~I $\lambda$6678 absorptions, assuming JD 2,452,695.5 for the
time of explosion. Note that the first H$\alpha$ expansion velocity
is an overestimate due to blending with Si~II $\lambda$6355; the same
may be true to a lesser extent for the second H$\alpha$ velocity.
(bottom) Expansion velocities for SN~2003bg (red),
SN~1987K [\cite{filippenko88}; green],
SN~1999em [\citet{hamuy01}, blue], 
and SN~1987A [\citet{phillips90}, magenta] 
measured from the minimum of the H$\alpha$ absorption, assuming 
JD 2,446,984 for the time of explosion of SN~1987K \citep{filippenko88},
JD 2,451,478.8 for the time of explosion of SN~1999em \citep{hamuy01},
and JD 2,446,849.82 for the time of explosion of SN~1987A \citep{svoboda87}.
}
\label{vel_fig}
\end{figure}

The expansion velocities for SN~2003bg during the optically thick
phase measured from the minimum of the H$\alpha$, H$\beta$, H$\gamma$,
He~I $\lambda$5876, and He~I $\lambda$6678 absorption lines are shown 
in the top panel of Figure \ref{vel_fig}.
We observe a rapid velocity drop during the first 20 days for the
Balmer lines, followed by a leveling off between days 20 and 50,
with H$\alpha$ exhibiting much higher velocities than the other two
Balmer lines throughout the photospheric phase. Both He~I lines have
considerably lower velocities than the H lines.  The higher velocity
of H$\alpha$ is the result of saturation in that line. The weaker
Balmer lines level off at a velocity of $\sim$10,000 km~s$^{-1}$,
indicating that H is confined to velocities higher than roughly this
value in the ejecta. The He lines level off at a lower velocity,
$\sim$7000 km~s$^{-1}$, which again indicates confinement of He to
velocities higher than this value. The stratification of both H and He
is also confirmed by spectral modeling \citep{mazzali09}.

Relative to most SNe~II, SN~2003bg stands out for its high
expansion velocity.  This can be clearly seen in the bottom of Figure
\ref{vel_fig}, which compares the H$\alpha$ velocity curve of
SN~2003bg (red), 
the Type IIb SN~1987K [\cite{filippenko88}; green],
the normal Type II-P SN~1999em [\citet{hamuy01}; blue], and 
SN~1987A [\cite{phillips90}; magenta].  
However, note that the initial H$\alpha$ expansion velocity of the first known
SN~IIb, SN 1987K, was also very high: \citet{filippenko88}
measured $v = 15,400$ km s$^{-1}$ in his first spectrum, which
was obtained near maximum brightness. This value is comparable to,
or even slightly exceeds, the H$\alpha$ expansion velocity of
SN 2003bg at a similar phase (March 12 spectrum, $v = 14,200$
km s$^{-1}$). Also, the velocity of the H$\alpha$ absorption minimum 
in the spectrum of SN 1987K steadily decreased with time, as in the
case of SN 2003bg. Had there been pre-maximum spectra of SN 1987K,
they may well have shown H$\alpha$ expansion velocities approaching
$\sim$20,000 km s$^{-1}$, close to that of SN 2003bg in the March      
2 spectrum.

The rapid expansion of SN~2003bg suggests that the kinetic energy per
unit mass is high. When the bolometric light curve (see Sec. 3.3) is
taken into account, the models of \citet{mazzali09} show that the
explosion kinetic energy is also very large. We conclude that (1)
broad-lined spectra are not exclusive to SNe in which hydrogen is
absent, and (2) SN 2003bg is the first known SN~II ``hypernova.''

\subsection{Spectroscopic Analysis During the Nebular Phase}

Our series of spectra obtained during the nebular phase is shown in
Figure \ref{spec1b_fig} and covers the period 176--301 days after
explosion. During this timespan the evolution of SN 2003bg was
slow. The most prominent features are forbidden lines such as the
[Ca~II] $\lambda\lambda$7292, 7324 and [O~I] $\lambda\lambda$6300,
6363 blends, with a FWHM of $\sim$5000 km~s$^{-1}$.  Weaker emissions
due to [S~II], Mg~I], [Fe~II], and [C~I] are also present. There is no
  clear evidence for broad H$\alpha$ emission from the SN. These
  late-time spectra are typical of SNe~Ib/c such as SN~1987M
  \citep{filippenko90,filippenko97}. Thus, SN~2003bg is a transitional
  SN~IIb, similar to SN~1993J \citep{filippenko93,filippenko94},
  although with much greater initial expansion velocities than the
  latter.

\subsection{Absolute Magnitudes and Bolometric Light Curve}

In this section we derive intrinsic properties of SN~2003bg such as
absolute magnitudes and the bolometric light curve. The first step
involves correcting the observed magnitudes for Galactic extinction,
for which we adopt $A_V^{\rm Gal} = 0.073$ mag \citep{schlegel98} and
the standard reddening law ($R_V = 3.1$) given by \citet{cardelli89}.

\begin{figure}
\plotone{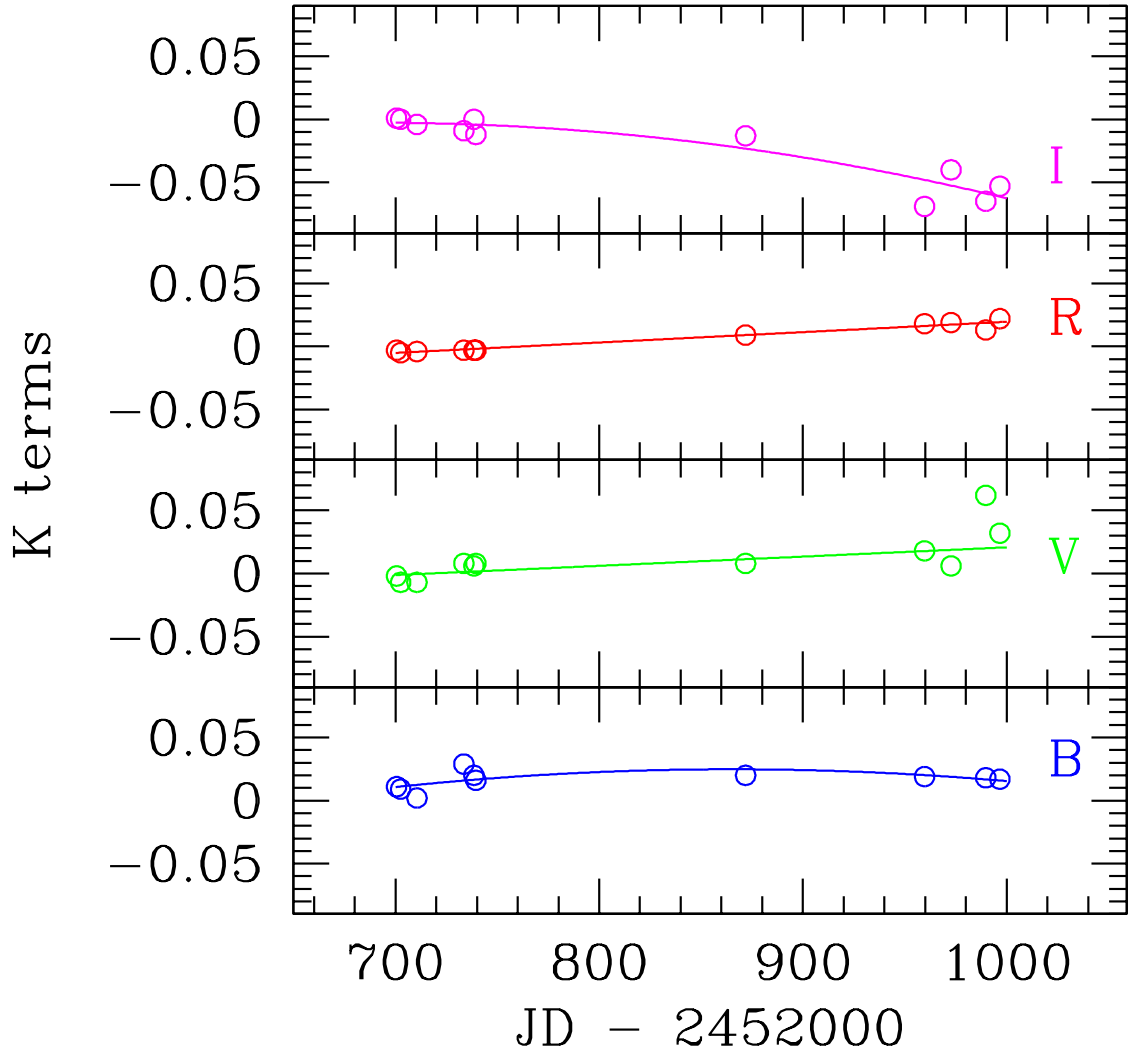}
\caption{K-terms in the $BVRI$ bands calculated from the spectra of
  SN~2003bg for $z = 0.00456$, as a function of Julian Day.
}
\label{kcorr_fig}
\end{figure}

The second step corrects the magnitudes for redshift-related effects
(K-terms). We use the definition appropriate for photon-counting
systems given by \citet{schneider83}, our spectra corrected for
Galactic extinction (excluding the data obtained with the Modular
Spectrograph), and the $BVRI$ filter passbands described by
\citet{bessell90}. The resulting K-terms for $cz = 1367$ km~s$^{-1}$
are shown in Figure \ref{kcorr_fig} as a function of JD. Given the
small redshift, the K-terms are small and exhibit a slow variation
with time. We use the cubic polynomial fits shown in this figure to
interpolate the K-terms to the time of our optical
photometry. The lack of IR spectra, on the other hand, prevents us
from K-correcting our $JHK$ magnitudes of SN~2003bg. We believe the
error introduced must be small owing to the small redshift of the SN.

The third step involves correcting the magnitudes for host-galaxy
extinction.  Examination of our spectra reveals no presence of
interstellar absorption lines of Na~I~D and Ca~II H \& K at the
redshift of the galaxy, suggesting no significant extinction. Thus, we
neglected such correction. The fourth and last step is to assume a
distance modulus, for which we adopt $\mu = 31.68 \pm 0.14$ mag
\citep{kelson00}.

\begin{figure}
\plotone{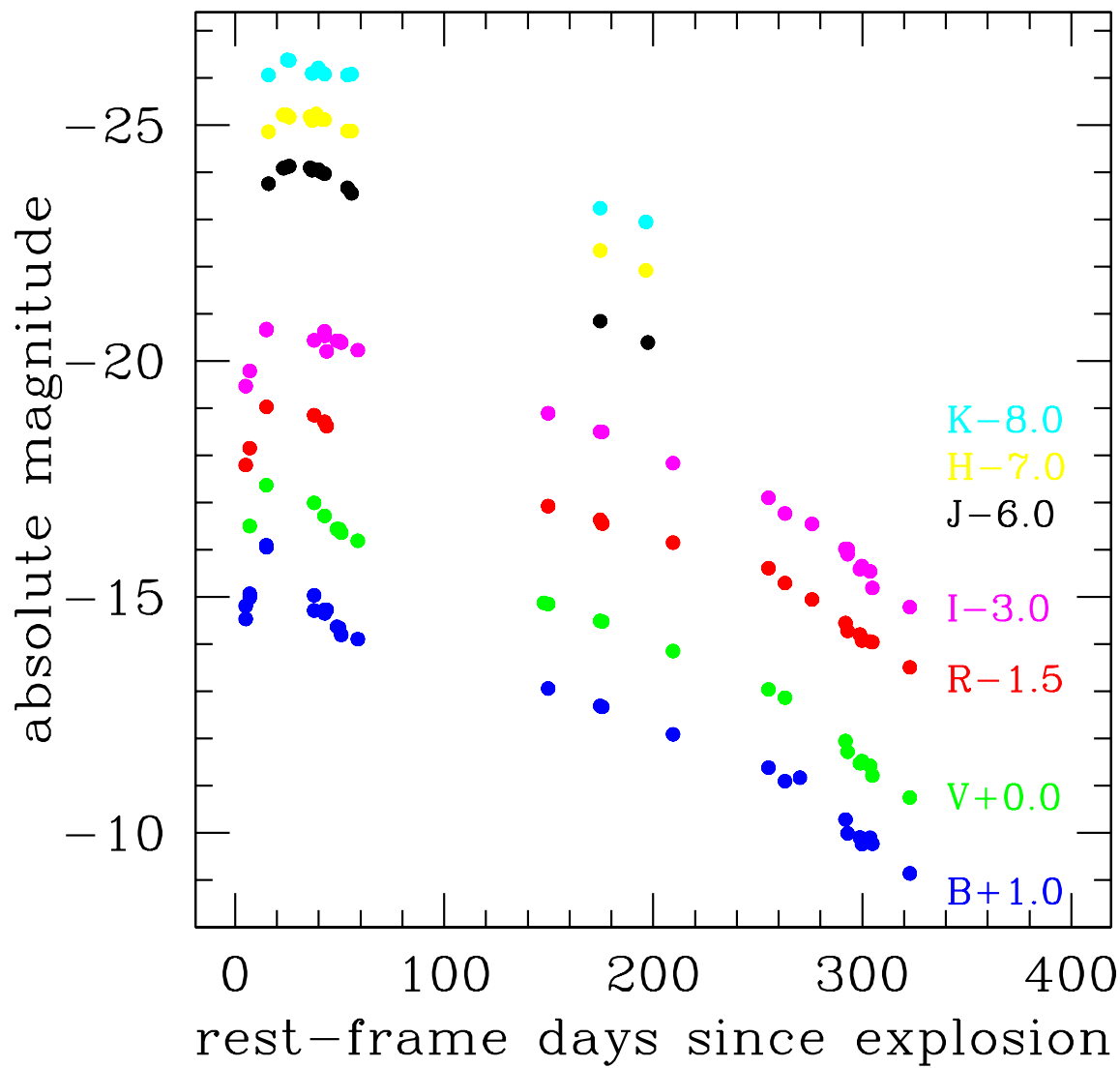}
\caption{$BVRIJ_sHK_s$ absolute magnitudes of SN~2003bg assuming
  $A_V^{\rm Gal} = 0.073$ mag, $A_V^{\rm host} = 0$ mag, 
  $d = 21.7$ Mpc, and an explosion
  time on JD 2,452,695.5. While the $BVRI$ were K-corrected using the
  K-terms shown in Figure \ref{kcorr_fig}, the $J_sHK_s$ magnitudes
  could not be K-corrected owing to the lack of IR spectra.
}
\label{absmag_fig}
\end{figure}

Figure \ref{absmag_fig} displays the resulting absolute magnitudes as a
function of rest-frame days since explosion (assumed to be on JD
2,452,695.5). The absolute peak $V$ magnitude of $M_V = -17.5$ does
not stand out compared with other core-collapse SNe. Since the light
curve is powered by the $^{56}$Ni $\rightarrow$ $^{56}$Co
$\rightarrow$ $^{56}$Fe decay chain, this luminosity implies that the
amount of Ni synthesized in the explosion was not particularly large,
despite the high expansion velocities shown by SN~2003bg.



\begin{figure}
\plotone{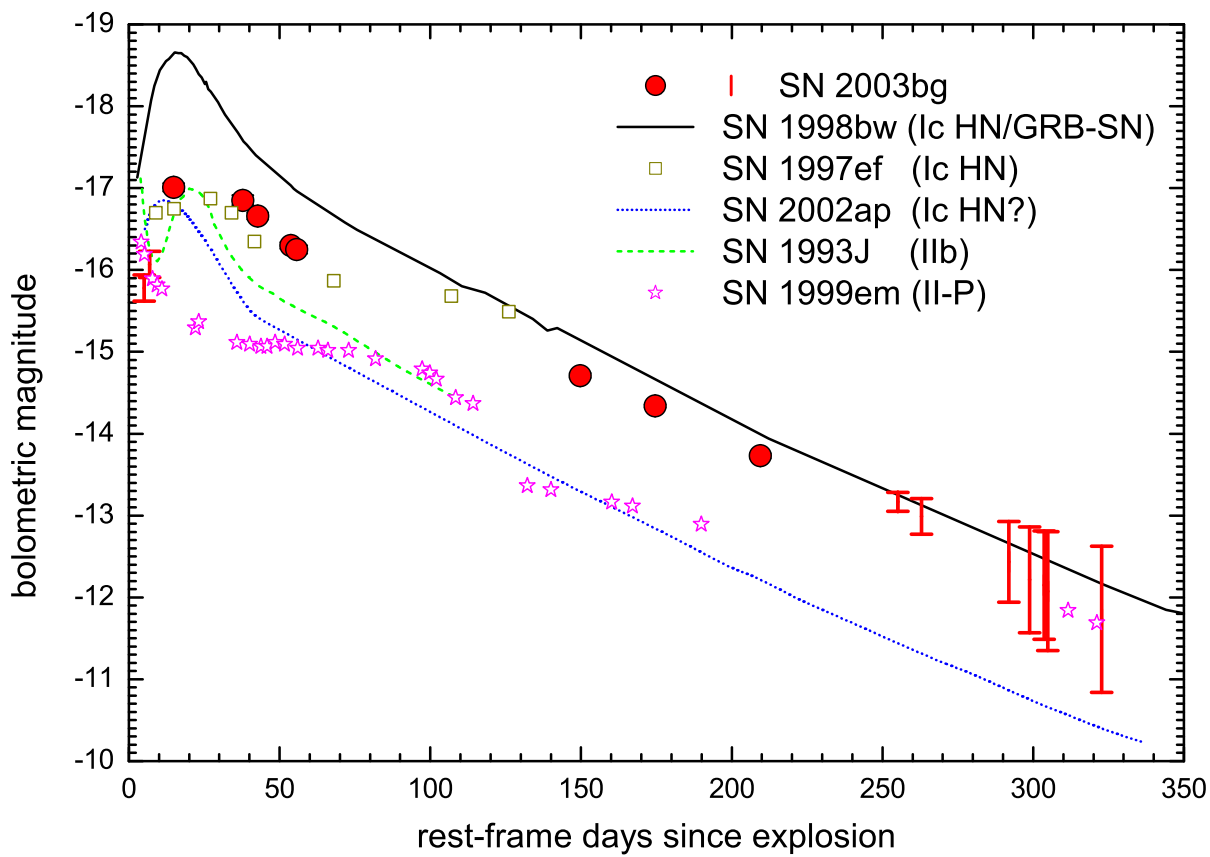}
\caption{
The $UVOIR$ bolometric light curve of SN~2003bg (red circles and
verticle bars), compared with those of the Type Ic SN~1998bw
[\citet{patat01}; black solid line] (a GRB-connected hypernova), Type
Ic SNe~1997ef [\citet{mazzali00,mazzali04}; dark yellow squares] and
2002ap [\citet{tomita06}; blue dotted line] (hypernovae without a
GRB connection), Type IIb SN~1993J [\citet{wada97}; green dashed line],
and Type II-P SN~1999em [\citet{elmhamdi03}; magenta stars].
}
\label{bolLC_fig}
\end{figure}

Bolometric magnitudes at day 15 and between days 37 and 210 (Figure
\ref{bolLC_fig}; circles) were calculated using our $BVRIJHK$
photometry.  A spectral energy distribution (SED) was built for each
epoch after converting the observed broad-band magnitudes into
monochromatic fluxes, using the conversion factors and effective
frequencies of \citet{bessell05}.  The SED was then integrated to
obtain the bolometric flux. No bolometric magnitude was estimated in
the $BVRI$ data gap (between days 15 and 37), where the light maximum
must lie. Before day 15, only the ranges of bolometric fluxes were
given (vertical lines). 

The lower limits were derived by
integrating $BVRI$ fluxes only, while the upper limits were found by
adding the same amount of NIR contribution as at day 15 ($\sim$
25\%). Both were then shifted by an extra $-0.1$ mag to account for
some $U$-band contribution which is possibly $\sim 10\%$ for a SN Ib/c
\citep{yoshii03}.  Since at late times the NIR flux percentage
increases as a SN evolves, the value at day 210 (34\%) was used
for any later epoch to get the lower bolometric limit. For the
late-time upper limits, we took the most conservative approach of
assuming a $^{56}$Co decay law from day 210 onward.
The resulting luminosities are listed in Table \ref{bol_tab}.



Figure \ref{bolLC_fig} shows the $UVOIR$ bolometric light curve of
SN~2003bg (red circles and verticle bars), compared with those of
other types of core-collapse SNe (II-P, IIb, Ic unrelated to GRB, and
GRB-related Ic).  Near maximum, SN~2003bg was nearly 1.5 dex fainter
than the GRB-connected Type Ic SN~1998bw [black solid line, 
\citet{patat01}], and similar in luminosity to the Type Ic
SNe~1997ef [dark yellow squares, \citet{mazzali00,mazzali04}] and
2002ap [blue dotted line, \citep{tomita06}], both of which were
hypernovae not associated with GRBs.  This suggests a normal (0.1--0.2
\Msun) Ni production in SN~2003bg, relative to other hypernovae not 
associated with GRBs. At late times, however, SN~2003bg appears as
luminous as SN~1998bw and much brighter than SN~2002ap. Relative to
all of these objects, SN~2003bg stands out for its broad light curve,
a possible indication of a larger than normal ejecta mass. Compared to
the non-hypernova Type IIb SN~1993J [green dashed line, \citet{wada97}], 
SN~2003bg was more luminous, especially at late times. The light curve 
is very different in shape and brighter than that of the typical 
Type II-P SN~1999em [magenta stars; \citet{elmhamdi03}].

\section{DISCUSSION AND CONCLUSIONS}
\label{disc}

Our early-time spectroscopic observations of SN~2003bg revealed the
presence of Balmer lines.  The nebular-phase spectra, on the other
hand, showed a remarkable spectroscopic resemblance to Type Ib/c
events, without clear evidence for hydrogen. These observations and
the corresponding modeling of \citet{mazzali09} indicate a progenitor
star with a hydrogen shell containing only $\sim$0.05 \Msun, similar
in mass to the Type IIb SN~1993J \citep{shigeyama94}. 

This picture is consistent with the radio model of
\citet{soderberg06}, which implies a dense circumstellar medium (CSM)
caused by the large mass-loss rate from the
progenitor. \citet{soderberg06} proposed that the bumps in the radio
light curve of SN 2003bg were due to a variable Wolf-Rayet (WR) wind,
but no physical mechanism for such variations was proposed. But
\citet{kotak06} instead suggested that these episodic bumps could be
due to a variable wind of a luminous blue variable star (LBV) with a
recurrence timescale of $\sim$25 years, qualitatively similar to the
modulations seen in the hypernova SN 1998bw.  The transitional
character of SN~2003bg as a Type IIb supernova reported here implies a
rather limited amount of H, consistent with an LBV scenario, as LBV
atmospheres are H-rich compared with WR stars, but He-rich compared to
OB stars and red supergiants \citep{kotak06}.

The expansion velocities inferred from the Balmer lines were initially
at the level of 20,000 km s$^{-1}$, though closer to 14,000 km s$^{-1}$ 
near maximum brightness; this is almost unprecedented for SNe~II
\citep[the notable exception being SN~IIb 1987K;][]{filippenko88}.
The first spectra bear resemblance to the broad-lined SNe~Ib and
SNe~Ic such as SN~1997ef. This suggests a larger than normal
kinetic energy per unit mass for a Type II event.  In fact, when the
bolometric light curve is taken into account, the total kinetic energy
of the ejecta is found to be large; the models of \citet{mazzali09}
yield $\sim 5 \times 10^{51}$ ergs, which is well above that of normal
core-collapse SNe ($\sim 10^{51}$ ergs) although definitely less than
that of the GRB-related SN~1998bw. Thus, SN~2003bg can be described as
the first known Type IIb hypernova. 

Near maximum brightness, SN~2003bg displayed a luminosity comparable
to that of other GRB-unrelated Type I hypernovae, implying a rather
normal amount of $^{56}$Ni production (0.1--0.2 \Msun).  The light
curve of SN~2003bg, on the other hand, is remarkably broad, thus
suggesting a relatively large progenitor mass at the moment of
explosion. In fact, the models in the accompanying paper by \citet{mazzali09}
find that SN~2003bg ejected $\sim$4~\Msun~ of material. This value, combined 
to the evolutionary models of \citet{nomoto88} suggests 
a ZAMS mass for the progenitor star of $\sim$20--25 \Msun. The uncertainty on 
these values depends primarily on the uncertain amount of mass loss
the star suffers during its evolution, which can affect the growth of the He
core. These conclusions are qualitatively
consistent with the mass-energy relation reported in the literature
for core-collapse SNe \citep[for SNe~II and SNe~Ib/c,
respectively]{hamuy03a,nomoto05}.

\acknowledgments

\noindent

M.H. acknowledges support provided by NASA through Hubble Fellowship
grant HST-HF-01139.01-A (awarded by the Space Telescope Science
Institute, which is operated by the Association of Universities for
Research in Astronomy, Inc., for NASA, under contract NAS 5-26555),
the Carnegie Postdoctoral Fellowship, FONDECYT through grant 1060808,
the Millennium Center for Supernova Science through grant P06-045-F
(funded by ``Programa Bicentenario de Ciencia y Tecnolog\'ia de
CONICYT'' and ``Programa Iniciativa Cient\'ifica Milenio de
MIDEPLAN''), Centro de Astrof\'\i sica FONDAP 15010003, and Center of
Excellence in Astrophysics and Associated Technologies (PFB 06).  This
research has been supported in part by National Natural Science
Foundation of China (Grant No. 10673014) and by National Basic
Research Program of China (Grant No.  2009CB824800).
A.V.F. gratefully acknowledges support from NSF grant AST--0607485 and
the TABASGO Foundation. Some of the data presented herein were
obtained at the W. M. Keck Observatory, which is operated as a
scientific partnership among the California Institute of Technology,
the University of California, and NASA; it was made possible by the
generous financial support of the W. M. Keck Foundation.  We wish to
extend special gratitude to those of Hawaiian ancestry on whose sacred
mountain we are privileged to be guests.


\clearpage

\begin{deluxetable} {lcccc}     
\tablecolumns{5}
\tablenum{1}
\tablewidth{0pc}
\tablecaption{Journal of Observations \label{journal_tab}}
\tablehead{
\colhead{Date (UT)} &
\colhead{Telescope} &
\colhead{Spectroscopy/} &
\colhead{Weather} &
\colhead{Observers\tablenotemark{a}}  \\
\colhead{} &
\colhead{} &
\colhead{Photometry?} &
\colhead{} &
\colhead{} 
} 
\startdata
2003 Feb 28 & Keck I     &  Spec  &    Clear\tablenotemark{b} & AF/RC \\
2003 Mar 02 & du Pont    &  Phot  &    Clear   & MR/NM \\
2003 Mar 02 & du Pont    &  Spec  &    Clear   & MR/NM \\
2003 Mar 03 & du Pont    &  Phot  &    Clear   & MP \\
2003 Mar 04 & du Pont    &  Phot  &    Clear   & MP \\
2003 Mar 04 & du Pont    &  Spec  &    Clear   & MP \\
2003 Mar 12 & du Pont    &  Phot  &    Clear   & MP/MR/NM \\
2003 Mar 12 & du Pont    &  Spec  &    Clear   & MP/MR/NM \\
2003 Mar 13 & du Pont    &  Phot  &    Cirrus  & MR \\
2003 Mar 20 & du Pont    &  Phot  &    Clear   & NM \\
2003 Mar 22 & du Pont    &  Phot  &    Clear   & NM \\
2003 Mar 23 & du Pont    &  Phot  &    Clouds  & NM \\
2003 Mar 31 & du Pont    &  Spec  &    Clear   & MH \\
2003 Apr 02 & Swope      &  Phot  &    Clear   & MH        \\
2003 Apr 03 & Swope      &  Phot  &    Clear   & MH        \\
2003 Apr 04 & du Pont    &  Phot  &    Clear   & JM/LH     \\
2003 Apr 04 & du Pont    &  Spec  &    Clear   & JM/LH \\ 
2003 Apr 05 & Swope      &  Phot  &    Clear   & MH        \\
2003 Apr 06 & Swope      &  Phot  &    Clear   & MH        \\
2003 Apr 07 & Swope      &  Phot  &    Clear   & MH        \\
2003 Apr 09 & Swope      &  Phot  &    Clear   & MH        \\
2003 Apr 09 & du Pont    &  Phot  &    Clear   & JM/LH     \\
2003 Apr 09 & du Pont    &  Spec  &    Clear   & JM/LH  \\
2003 Apr 10 & Clay       &  Phot  &    Clear   & MH \\
2003 Apr 10 & Clay       &  Spec  &    Clear   & MH \\
2003 Apr 15 & CTIO 0.9-m &  Phot  &    Cirrus  & LG \\
2003 Apr 16 & Swope      &  Phot  &    Cirrus  & SG \\
2003 Apr 16 & CTIO 0.9-m &  Phot  &    Cirrus? & LG \\
2003 Apr 17 & CTIO 0.9-m &  Phot  &    Cirrus? & LG \\
2003 Apr 20 & du Pont    &  Phot  &    Clear   & MP/NM \\
2003 Apr 22 & du Pont    &  Phot  &    Clear   & MP    \\
2003 Apr 25 & Swope      &  Phot  &    Cirrus  & JT-O/NM \\
2003 Jul 23 & Swope      &  Phot  &    Clouds  & LG \\
2003 Jul 25 & Swope      &  Phot  &    Clear   & SG/LG \\
2003 Aug 19 & du Pont    &  Phot  &    Clear   & NM \\
2003 Aug 19 & Swope      &  Phot  &    Clear   & KK \\
2003 Aug 20 & du Pont    &  Phot  &    Cirrus  & NM \\
2003 Aug 20 & du Pont    &  Spec  &    Cirrus  & NM \\
2003 Sep 10 & du Pont    &  Phot  &    Clear   & kk \\
2003 Sep 11 & du Pont    &  Phot  &    Clear   & kk \\
2003 Sep 18 & du Pont    &  Spec  &    Cirrus  & NM \\
2003 Sep 23 & Swope      &  Phot  &    Clear   & LG \\
2003 Nov 08 & Swope      &  Phot  &    Clear   & SG \\
2003 Nov 16 & Clay       &  Phot  &    Clouds  & NM \\
2003 Nov 16 & Clay       &  Spec  &    Clouds  & NM \\
2003 Nov 16 & Swope      &  Phot  &    Clouds  & KK \\
2003 Nov 23 & Swope      &  Phot  &    Cirrus  & LG \\
2003 Nov 29 & du Pont    &  Spec  &    Clear   & MR \\
2003 Dec 15 & Swope      &  Phot  &    Clear   & SG \\
2003 Dec 16 & du Pont    &  Phot  &    Clear   & MH \\
2003 Dec 16 & du Pont    &  Spec  &    Clear   & MH \\
2003 Dec 22 & Swope      &  Phot  &    Clear   & SG \\
2003 Dec 23 & du Pont    &  Phot  &    Clear   & NM \\
2003 Dec 23 & du Pont    &  Spec  &    Clear   & NM \\
2003 Dec 27 & Swope      &  Phot  &    Clear   & SG \\
2003 Dec 28 & Swope      &  Phot  &    Clear   & SG \\
2004 Jan 15 & Swope      &  Phot  &    Clear   & LG \\
2004 Nov 11 & du Pont    &  Phot  &    Cirrus  & NM \\
2005 Feb 13 & du Pont    &  Phot  &    Cirrus  & NM \\

\enddata
\tablenotetext{a} 
{RC: Ryan Chornock,
AF: Alex Filippenko,
LG: Luis Gonz\'alez, 
SG: Sergio Gonzalez, 
MH: Mario Hamuy, 
LH: Leonor Huerta, 
kk: Kathleen Koviak,
KK: Kevin Krisciunas,
JM: Jos\'e Maza,
NM: Nidia Morrell,
MP: Mark Phillips, 
MR: Miguel Roth,
JT-O: Joanna Thomas-Osip.}
\tablenotetext{b}{SN~2003bg was observed at very high airmass
and was partially vignetted by the dome shutter.}

\end{deluxetable}

\clearpage

\begin{deluxetable} {ccccc}
\tablecolumns{4}
\tablenum{2}
\tablewidth{0pc}
\tablecaption{$BRVI$ Photometric Sequence Around SN~2003bg \label{sequence_tab}}
\tablehead{
\colhead{Star} &
\colhead{$B$ (mag)} &
\colhead{$V$ (mag)} &
\colhead{$R$ (mag)} &
\colhead{$I$ (mag)}  }

\startdata

c1 &  16.623(013) &  16.051(006) &  15.708(009) &  15.366(005) \\
c2 &  16.472(012) &  15.632(007) &  15.106(008) &  14.637(005) \\
c3 &  17.937(014) &  16.891(008) &  16.238(011) &  15.702(005) \\
c4 &  18.276(015) &  17.707(008) &  17.349(010) &  17.019(010) \\
c5 &  19.841(012) &  18.386(007) &  17.443(012) &  16.496(006) \\
c6 &  18.615(014) &  17.927(008) &  17.530(009) &  17.128(009) \\
c7 &  19.756(013) &  19.310(013) &  18.960(017) &  18.539(020) \\
c8 &  21.175(050) &  19.745(025) &  18.821(010) &  17.927(014) \\
c9 &  17.661(009) &  16.685(009) &  16.090(009) &  15.596(005) \\
c10&  20.547(029) &  19.442(026) &  18.717(010) &  18.071(015) \\

\enddata
\tablecomments{Uncertainties given in parentheses in thousandths of a magnitude,
corresponding to the root-mean square of the magnitudes obtained on eight photometric 
nights, with an uncertainty of 0.015 mag for an individual measurement.}
\end{deluxetable}

\clearpage

\begin{deluxetable} {ccccc}
\tablecolumns{5}
\tablenum{3}
\tablewidth{0pc}
\tablecaption{Color Terms for the Four Optical Cameras \label{coefficients_tab}}
\tablehead{
\colhead{Telescope} &
\colhead{$B$} &
\colhead{$V$} &
\colhead{$R$} &
\colhead{$I$}  
}

\startdata

Swope/CCD           & +0.053  & $-$0.054 & +0.021  & +0.052  \\
du Pont/WFCCD       & +0.125  & $-$0.045 & \nodata & +0.010  \\
Clay/LDDS-2         & +0.132  & +0.046 & \nodata & \nodata \\
CTIO 0.9-m/CCD      & $-$0.086  & +0.011 & +0.004  & +0.007  \\

\enddata
\tablecomments{Color terms are defined in equations \ref{B_eq}--\ref{I_eq}.}
\end{deluxetable}

\clearpage

\begin{deluxetable} {cccccccc}
\rotate
\tablecolumns{8}
\tablenum{4}
\tablewidth{0pc}
\tablecaption{$BV(RI)_{KC}$ Photometry for SN~2003bg \label{photometry_tab}}
\tablehead{
\colhead{Date (UT)} &
\colhead{JD$-$2,400,000} &
\colhead{$B$ (mag)} &
\colhead{$V$ (mag)} &
\colhead{$R$ (mag)} &
\colhead{$I$ (mag)} &
\colhead{Method} &
\colhead{Telescope} }
\startdata

2003 Mar 02 & 52700.6 & 15.979(017) &  \nodata     &  \nodata     &  \nodata     & DAO &  du Pont \\ 
2003 Mar 02 & 52700.6 & 16.252(100) &  \nodata     &  15.436(100) &  15.252(100) & SYN &  du Pont \\
2003 Mar 04 & 52702.5 & 15.720(024) &  15.244(012) &  \nodata     &  \nodata     & DAO &  du Pont \\
2003 Mar 04 & 52702.5 & 15.803(100) &  \nodata     &  15.077(100) &  14.930(100) & SYN &  du Pont \\
2003 Mar 12 & 52710.5 & 14.692(017) &  14.380(023) &  \nodata     &  14.062(028) & DAO &  du Pont \\
2003 Mar 12 & 52710.6 & 14.735(100) &  \nodata     &  14.204(100) &  14.041(100) & SYN &  du Pont \\
2003 Apr 04 & 52733.5 & 15.751(012) &  14.756(018) &  \nodata     &  \nodata     & DAO &  du Pont \\
2003 Apr 04 & 52733.5 & 16.071(100) &  \nodata     &  14.382(100) &  14.277(100) & SYN &  du Pont \\
2003 Apr 09 & 52738.5 & 16.132(028) &  15.033(025) &  \nodata     &  14.082(031) & DAO &  du Pont \\
2003 Apr 09 & 52738.5 & 16.061(100) &  \nodata     &  14.523(100) &  14.179(100) & SYN &  du Pont \\
2003 Apr 10 & 52739.5 & 16.059(100) &  \nodata     &  14.614(100) &  14.508(100) & SYN &  Clay   \\
2003 Apr 15 & 52744.5 & 16.418(016) &  15.312(014) &  \nodata     &  14.282(014) & DAO &  CTIO 0.9-m \\
2003 Apr 16 & 52745.5 & \nodata     &  15.315(014) &  \nodata     &  14.292(014) & DAO &  CTIO 0.9-m \\
2003 Apr 16 & 52745.5 & 16.445(016) &  15.322(014) &  \nodata     &  14.287(014) & DAO &  Swope \\
2003 Apr 17 & 52746.5 & 16.590(040) &  15.389(014) &  \nodata     &  14.320(014) & DAO &  CTIO 0.9-m \\
2003 Apr 25 & 52754.5 & 16.679(016) &  15.563(014) &  \nodata     &  14.481(014) & DAO &  Swope \\
2003 Jul 23 & 52843.9 & \nodata     &  16.888(014) &  \nodata     &  \nodata     & DAO &  Swope \\
2003 Jul 25 & 52845.9 & 17.730(017) &  16.914(014) &  16.319(015) &  15.800(014) & DAO &  Swope \\
2003 Aug 19 & 52870.9 & 18.098(016) &  17.276(016) &  16.611(018) &  16.184(014) & DAO &  Swope \\
2003 Aug 20 & 52871.9 & \nodata     &  17.287(025) &  \nodata     &  \nodata     & DAO &  du Pont \\
2003 Aug 20 & 52871.9 & 18.121(100) &  \nodata     &  16.691(100) &  16.191(100) & SYN &  du Pont \\
2003 Sep 23 & 52905.9 & 18.700(016) &  17.918(015) &  17.097(022) &  16.846(016) & DAO &  Swope \\
2003 Nov 08 & 52951.8 & 19.409(041) &  18.737(030) &  17.643(018) &  17.571(023) & DAO &  Swope \\
2003 Nov 16 & 52959.7 & \nodata     &  18.911(044) &  \nodata     &  \nodata     & DAO &  Swope \\
2003 Nov 16 & 52959.7 & 19.695(100) &  \nodata     &  17.960(100) &  17.901(100) & SYN &  Clay   \\
2003 Nov 23 & 52966.9 & 19.623(107) &  \nodata     &  \nodata     &  \nodata     & DAO &  Swope \\
2003 Nov 29 & 52972.7 & \nodata     &  \nodata     &  18.309(100) &  18.121(100) & SYN &  du Pont \\
2003 Dec 15 & 52988.8 & 20.509(057) &  19.835(052) &  18.809(024) &  18.647(040) & DAO &  Swope \\
2003 Dec 16 & 52989.8 & \nodata     &  20.062(076) &  \nodata     &  18.648(044) & DAO &  du Pont \\
2003 Dec 16 & 52989.8 & 20.803(100) &  \nodata     &  18.976(100) &  18.758(100) & SYN &  du Pont \\
2003 Dec 22 & 52995.7 & 20.890(061) &  20.307(060) &  19.054(028) &  19.075(058) & DAO &  Swope \\
2003 Dec 23 & 52996.7 & \nodata     &  20.262(049) &  \nodata     &  \nodata     & DAO &  du Pont \\
2003 Dec 23 & 52996.7 & 21.030(100) &  \nodata     &  19.179(100) &  19.009(100) & SYN &  du Pont \\
2003 Dec 27 & 53000.6 & 20.894(060) &  20.361(056) &  19.198(025) &  19.122(044) & DAO &  Swope \\
2003 Dec 28 & 53001.7 & 21.023(048) &  20.567(064) &  19.214(032) &  19.472(068) & DAO &  Swope \\
2004 Jan 15 & 53019.7 & 21.652(079) &  21.036(070) &  19.750(020) &  19.876(048) & DAO &  Swope \\

\enddata
\tablecomments{DAO and SYN mean PSF and synthetic photometry, respectively;
uncertainties are given in parentheses in thousandths of a magnitude.}
\end{deluxetable}

\clearpage

\begin{deluxetable} {cccccc}
\tablecolumns{6}
\tablenum{5}
\tablewidth{0pc}
\tablecaption{$J_sHK_s$ Photometry for SN~2003bg \label{ir_photometry_tab}}
\tablehead{
\colhead{Date (UT)} &
\colhead{JD$-$2,400,000} &
\colhead{$J_s$ (mag)} &
\colhead{$H$ (mag)} &
\colhead{$K_s$ (mag)} &
\colhead{Telescope} }
\startdata

2003 Mar 13 & 52711.56 &   13.936(023) &  13.827(030) &  13.622(045) &  du Pont \\
2003 Mar 20 & 52718.59 &   13.609(029) &  13.474(033) &  \nodata      &  du Pont \\
2003 Mar 22 & 52720.53 &   13.584(023) &  13.477(033) &  13.301(043) &  du Pont \\
2003 Mar 23 & 52721.55 &   13.565(025) &  13.520(039) &  13.312(042) &  du Pont \\
2003 Apr 02 & 52731.55 &   13.601(026) &  13.510(051) &  \nodata      &  Swope  \\
2003 Apr 03 & 52732.52 &   13.650(028) &  13.592(044) &  13.588(089) &  Swope  \\
2003 Apr 05 & 52734.52 &   13.638(025) &  13.451(039) &  \nodata      &  Swope  \\
2003 Apr 06 & 52735.52 &   13.641(023) &  \nodata     &  13.471(104) &  Swope  \\
2003 Apr 07 & 52736.52 &   13.679(025) &  13.557(046) &  \nodata      &  Swope  \\
2003 Apr 09 & 52738.50 &   13.724(024) &  13.573(036) &  13.603(083) &  Swope  \\
2003 Apr 20 & 52749.50 &   14.025(026) &  13.812(034) &  13.623(057) &  du Pont \\
2003 Apr 22 & 52751.48 &   14.141(029) &  13.812(040) &  13.603(059) &  du Pont \\
2003 Aug 19 & 52870.92 &   16.850(036) &  16.346(040) &  16.443(110) &  du Pont \\
2003 Sep 10 & 52892.90 &   \nodata     &  16.768(047) &  16.736(092) &  du Pont \\
2003 Sep 11 & 52893.82 &   17.302(028) &  \nodata     &  \nodata      &  du Pont \\

\enddata
\tablecomments{Uncertainties given in parentheses in thousandths of a magnitude.}
\end{deluxetable}

\clearpage

\begin{deluxetable} {ccccccc}
\tablecolumns{7}
\tablenum{6}
\tablewidth{0pc}
\tablecaption{Spectroscopic Observations of SN~2003bg \label{optspec_tab}}
\tablehead{
\colhead{Date (UT)} &
\colhead{JD} &
\colhead{$t-t_0$\tablenotemark{a}} &
\colhead{Instrument} &
\colhead{Wavelength} &
\colhead{Resolution} &
\colhead{Exposure} \\
\colhead{} &
\colhead{$-$2,400,000} &
\colhead{(days)} &
\colhead{} &
\colhead{Range (\AA)} &
\colhead{(\AA)} &
\colhead{(s)} }
\startdata

2003 Feb 28 & 52698.83 &  3.3    & LRIS      & 3200--9420 & 6  &  180 \\
2003 Mar 02 & 52700.58 &  5.0    & WFCCD     & 3800--9330 & 6  & 1200 \\
2003 Mar 04 & 52702.54 &  7.0    & WFCCD     & 3800--9330 & 6  & 1200 \\
2003 Mar 12 & 52710.55 & 15.1    & WFCCD     & 3800--9330 & 6  & 1200 \\
2003 Mar 31 & 52729.53 & 34.0    & Mod. Spec.& 3780--7280 & 7  &  600 \\
2003 Apr 04 & 52733.54 & 38.0    & WFCCD     & 3800--9330 & 6  &  600 \\
2003 Apr 09 & 52738.53 & 43.0    & WFCCD     & 3800--9330 & 6  &  300 \\
2003 Apr 10 & 52739.49 & 44.0    & LDSS-2    & 3600--9000 & 14 &  300 \\
2003 Aug 20 & 52871.89 &176.4    & WFCCD     & 3800--9330 & 6  &  900 \\
2003 Sep 18 & 52900.90 &205.4    & Mod. Spec.& 3780--7280 & 7  &  900 \\
2003 Nov 16 & 52959.74 &264.2    & LDSS-2    & 3600--9000 & 14 &  900 \\
2003 Nov 29 & 52972.73 &277.2    & WFCCD     & 4040--9330 & 6  & 1800 \\
2003 Dec 16 & 52989.79 &294.3    & WFCCD     & 3800--9330 & 6  & 1800 \\
2003 Dec 23 & 52996.72 &301.2    & WFCCD     & 3800--9330 & 6  & 1800 \\

\enddata
\tablenotetext{a}{Time of explosion assumed to be on JD 2,452,695.5.}
\end{deluxetable}

\clearpage

\begin{deluxetable} {ccccc}
\tablecolumns{5}
\tablenum{7}
\tablewidth{0pc}
\tablecaption{Bolometric Magnitudes of SN~2003bg \label{bol_tab}}
\tablehead{
\colhead{$t-t_0$} &
\colhead{Bolometric} &
\colhead{Error} &
\colhead{Middle of Upper} &
\colhead{Error} \\
\colhead{(days)} &
\colhead{Magnitude} &
\colhead{} &
\colhead{\& Lower Limit} &
\colhead{} }
\startdata

5.05   & \nodata & \nodata&  $-$15.78    &  0.16   \\
6.94   & \nodata & \nodata&  $-$16.07    &  0.16   \\
14.91  & $-$17.01  & 0.05   &   \nodata  &  \nodata\\
37.80  & $-$16.85  & 0.06   &   \nodata  &  \nodata\\
42.78  & $-$16.66  & 0.04   &   \nodata  &  \nodata\\
53.70  & $-$16.30  & 0.04   &   \nodata  &  \nodata\\
55.70  & $-$16.25  & 0.05   &   \nodata  &  \nodata\\
149.69 & $-$14.71  & 0.03   &   \nodata  &  \nodata\\
174.58 & $-$14.34  & 0.02   &   \nodata  &  \nodata\\
209.42 & $-$13.73  & 0.03   &   \nodata  &  \nodata\\
255.11 & \nodata & \nodata&  $-$13.17    &  0.12   \\
262.98 & \nodata & \nodata&  $-$12.99    &  0.22   \\
291.94 & \nodata & \nodata&  $-$12.43    &  0.49   \\
298.81 & \nodata & \nodata&  $-$12.21    &  0.64   \\
303.69 & \nodata & \nodata&  $-$12.15    &  0.66   \\
304.79 & \nodata & \nodata&  $-$12.08    &  0.73   \\
322.70 & \nodata & \nodata&  $-$11.73    &  0.89   \\

\enddata
\tablecomments{Adopted values: \\
$E(B-V)^{\rm Gal} = 0.02$ mag, $E(B-V)^{\rm host} = 0.00$ mag; \\
$d = 21.7$ Mpc; \\
$t_0 =$ JD 2,452,695.5.}
\end{deluxetable}


%


\clearpage

\clearpage

\clearpage

\clearpage

\clearpage

\clearpage

\clearpage

\clearpage

\clearpage

\clearpage

\clearpage

\clearpage

\end{document}